\documentclass[12pt]{amsart}
\usepackage{amsbsy,amssymb,amsmath,amsthm,amscd,amsfonts,latexsym,amstext,delarray,
amsmath,graphicx} 
\usepackage{qtree}
\usepackage[margin=.8in]{geometry}
\usepackage{color}
\usepackage[all]{xy}

\newtheorem{thm}{Theorem}[section]
\newtheorem{prop}{Proposition}[section]

\newtheorem{lem}[thm]{Lemma}

\newtheorem{defn}[prop]{Definition}
\newtheorem{rem}[prop]{Remark}

\usepackage[normalem]{ulem}

\usepackage[]{mdframed}

\numberwithin{equation}{section}

\def\cA{{\mathcal A}}
\def\cB{{\mathcal B}}
\def\cC{{\mathcal C}}

\def\cE{{\mathcal E}}
\def\cF{{\mathcal F}}
\def\cG{{\mathcal G}}

\def\cI{{\mathcal I}}

\def\cL{{\mathcal L}}
\def\cM{{\mathcal M}}

\def\cO{{\mathcal O}}
\def\cP{{\mathcal P}}
\def\cQ{{\mathcal Q}}
\def\cR{{\mathcal R}}
\def\cS{{\mathcal S}}
\def\cT{{\mathcal T}}

\def\cW{{\mathcal W}}
\def\cX{{\mathcal X}}

\def\bB{{\mathbb B}}

\def\bS{{\mathbb S}}
\def\bT{{\mathbb T}}

\def\Hom{{\rm Hom}} 

\def\fA{{\mathfrak A}}
\def\fB{{\mathfrak B}}
\def\fC{{\mathfrak C}}

\def\fM{{\mathfrak M}}
\def\fN{{\mathfrak N}}

\def\fB{{\mathfrak B}}

\def\fQ{{\mathfrak Q}}

\def\fN{{\mathfrak N}}
\def\fc{{\mathfrak c}}

\title{Formal Languages and TQFTs with Defects}
\author{Luisa Boateng and Matilde Marcolli}
\date{2024}

\address{Department of Mathematics, Stanford University, Stanford, CA 94305, USA}
\email{luisa.mboateng@gmail.com}
\address{Department of Mathematics and Department of Computing and Mathematical Sciences, 
California Institute of Technology, CA 91125, USA}
\email{matilde@caltech.edu}

\begin{document}
\maketitle

\begin{abstract}
A construction that assigns a Boolean 1D TQFT with defects to a finite state automaton was recently
developed by Gustafson, Im, Kaldawy, Khovanov, and Lihn. We show that the construction
is functorial with respect to the category of finite state automata with transducers as morphisms.
Certain classes of subregular languages correspond to additional
cohomological structures on the associated TQFTs.
We also show that the construction generalizes to context-free grammars through a categorical version of
the Chomsky--Sch\"utzenberger representation theorem, due to Melli\`es and Zeilberger. 
The corresponding TQFTs are then described as morphisms of colored operads on
an operad of cobordisms with defects. 
 \end{abstract}

\section{Introduction}

A recent series of papers \cite{ImKho}, \cite{GuImKaKhoLi}, \cite{ImKho2}, \cite{GuImKho} showed
that nondeterministic finite state automata (which correspond to regular languages in the
theory of formal languages) give rise to one-dimensional oriented Topological Quantum Field
Theories (TQFTs) with zero-dimensional defects (corresponding to the letters of the
formal language alphabet), so that the regular language computed by the automaton 
has a path integral interpretation. 
This approach, seen from both the formal languages perspective and the TQFT side, 
suggests several possible questions. 

\smallskip

Regular languages, which are computed by finite state automata, are the most restrictive
class of the Chomsky hierarchy \cite{Chomsky59} of formal languages. The next larger class
in the hierarchy of formal languages is the context-free languages, which are computed by
pushdown stack automata. While these form a very large class of formal languages, a well
known theorem of Chomsky--Sch\"utzenberger \cite{ChoSchu} reduces all context-free languages to
a combination (realized by an intersection and a homomorphism) of a regular language
and one particular context free language called the Dyck language (see also \cite{Kamb} 
for a recent survey of this result). This suggests that a physical interpretation of the Dyck
languages would suffice, in combination with
the 1-dimensional TQFTs associated to regular languages, to extend the construction of
physical systems to arbitrary context-free grammars. 

\smallskip

In this paper we consider this question from a category-theoretic perspective,
on the basis of a recent categorical reformulation of
the Chomsky--Sch\"utzenberger theorem, developed in \cite{MelZei}, \cite{MelZei2}, and and
additional operadic structure on the side of one-dimensional TQFTs with defects.

\smallskip

We first discuss the category of regular languages (or of context-free languages) with
morphisms given by rational transductions, as described in \cite{FerMar} and
the associated category of finite state automata (or of pushdown stack automata)
with transducers as morphisms. We show that the construction of 1D Boolean
TQFTs with defects of \cite{GuImKaKhoLi} associated to finite state automata
is functorial with respect to transducers. 

\smallskip

We then consider a category-theoretic version of finite state automata introduced in
\cite{MelZei} and we show that the construction of \cite{GuImKaKhoLi} of 
associated 1D Boolean TQFTs with defects extends to categorical finite state automata
with categorical transducers as morphisms. 

\smallskip

We then use the categorical version of context-free grammars introduced in 
\cite{MelZei}, \cite{MelZei2}, based on operads of sliced arrows in a category,
with morphisms given by categorical transducers. 
We show that a functorial assignment of associated 1D Boolean TQFTs with defects
can be obtained in this case too, using an operad of cobordisms with defects.
We show that the categorical version of the Chomsky--Sch\"utzenberger 
representation theorem proved  \cite{MelZei2} implies that this construction
of 1D Boolean TQFTs with defects is entirely determined by its value on
certain tree-contour grammars (the categorical version of the Dyck languages)
and by functoriality under transducers. 

\smallskip

\section{Categories of Formal Languages and Automata}\label{AutoCatSec}

\subsection{A category of formal languages} \label{CatFLsec}

We recall here the basic definitions on regular and context-free formal languages
that we will be using in the rest of this section. We also review the construction of
\cite{FerMar} of a category of formal languages, by distinguishing the regular and
the context-free case, and the cases of automata versus languages. 

\smallskip
\subsubsection{Regular and context-free languages} 

\begin{defn}\label{FSAPSA}
Let $\fA$ be a finite alphabet. 
A finite state automaton (FSA) is a set  $\fM=(\fQ,\fA,\tau,q_0, F)$ with a finite set $\fQ$
of possible states, a subset $F\subset \fQ$ of final states and an initial state $q_0$, and
a set $\tau\subset \fQ\times \fA \times \fQ$ of transitions. A pushdown stack automaton (PSA) 
is a set $\fM=(\fQ, \fA, \Gamma, \tau, q_0, F, z_0)$ with $\fQ,F,\fA,q_0$ as in the FSA case
and with $\Gamma$ the stack alphabet and set of transitions 
$\tau\subset \fQ\times (\fA\cup \{ \epsilon \})\times \Gamma \times \fQ \times \Gamma^*$ 
and with an initial symbol $z_0\in \Gamma$ for the stack. We do not assume that
the automata we consider are deterministic. 
\end{defn}

Equivalently a FSA $\fM$ is a finite directed graph with $\fQ$ as the set of vertices and
a directed edge labelled by $a$ from $q$ to $q'$ for every transition of the form $(q,a,q')$.
This is interpreted as the automaton moving from state $q$ to state $q'$ upon reading the
letter $a$ on an instruction tape. 
A PSA has a similar directed graph, together with a stack memory that is only accessible last-in-first-out,
so the transitions  $(q,a,z,q',\gamma)$ correspond to reading a
letter $a\in \fA$ on an instruction tape and at the same time reading a letter $z\in \Gamma$ at the top
of the stack, causing the automaton to move from state $q$ to state $q'$ and deposit a new 
string $\gamma\in \Gamma^*$ at the top of the stack. 

\smallskip

For a given finite alphabet $\fA$, one can form the free monoid $\fA^*$ generated by $\fA$.
A formal language on the alphabet $\fA$ is a subset $\cL\subset \fA^*$. A grammar for
a formal language $\cL$ is a datum $\cG=( \fN, \fA, P, S)$ consisting of a set $\fN$ of
{\em nonterminal symbols}, the alphabet $\fA$ (the set of {\em terminal symbols}), a start symbol $S$,
and a finite set of {\em production rules} $P$, with the property the words $w\in \cL$ are all
the strings of terminal symbols obtainable through a finite sequence of applications of
production rules in $P$ starting from $S$. Regular and context-free languages are described as:
\begin{itemize}
\item $\cL$ is a regular language if it can be produced by a grammar $G$ where
all the production rules are of the form $R \to a$ or $R \to Ra$, for some $R\in \fN$ 
and some $a\in \fA$.
\item  $\cL$ is a context-free language if it can be produced by a grammar $G$ where
the production rules are of the form $R \to \alpha$ where $\alpha$ is a string of terminals 
and non-terminals. 
\end{itemize}
The main classification theorem for formal languages is the Chomsky hierarchy 
 \cite{Chomsky56}, \cite{Chomsky59}, that identifies four increasingly more general 
 classes of formal languages in terms of their computability by different classes
 of automata. In particular, regular languages are computed by FSAs and
 context-free languages are computed by PSAs.

\smallskip
\subsubsection{Category of formal languages}\label{catFLsec}

We recall from  \cite{FerMar} the construction of a category whose objects are formal
languages (regular and context-free) and whose morphisms are 
correspondences defined by transductions.

\smallskip

For $\cL\subset \fB^*$ and $\cL'\subset \fC^*$ (regular or 
context-free) languages in the respective alphabets $\fB$ and $\fC$, 
a {\em rational transduction} $X=X(\fA,\alpha,\beta,\cL_{reg})$,
with $\alpha: \fA^* \to \fB^*$ and $\beta: \fA^* \to \fC^*$, and with $\cL_{reg} \subset \fA^*$
a regular language, is given by
$$ X =\{ (\alpha(w), \beta(w)) \,|\, w\in \cL_{reg} \} \subset \fB^* \times \fC^* $$
with $\cL \subset \alpha(\fA^*)$ and $\cL'\subset \beta(\fA^*)$, and such that $X(\cL)\subset\cL'$.
Here for a subset $\Omega \subset \alpha(\fA^*)\subset \fB^*$ one writes
$X(\Omega):=\beta(\alpha^{-1}(\Omega)\cap \cL_{reg}) \subset \fC^*$.

\begin{defn}\label{catFL}
The category $\cF\cL$ has objects that are formal languages $\cL$ that are
either regular or context-free, and non-identity morphisms given by rational
transductions $X \in {\rm Mor}_{\cF\cL}(\cL,\cL')$, for $\cL\subset \fB^*$ and
$\cL'\subset \fC^*$, with the composition law
\begin{equation}\label{corrcompose}
 Y\circ X=((\fA_1\cup \fA_2)^*, \alpha\circ \pi_{\fA_1}, \beta' \circ \pi_{\fA_2}, \hat\cL) \, , 
\end{equation} 
for $X=X(\fA_1,\alpha,\beta,\cL)$ and $Y=Y(\fA_2, \alpha',\beta', \cL')$, 
where $\hat\cL \subset \pi_{\fA_1}^{-1}(\cL) \cap \pi_{\fA_2}^{-1}(\cL')$ given by 
\begin{equation}\label{hatL}
\hat\cL=\{ w\in \pi_{\fA_1}^{-1}(\cL)\cap \pi_{\fA_2}^{-1}(\cL') \,|\, \beta\circ \pi_{\fA_1}(w) = \alpha'\circ\pi_{\fA_2}(w) \}\, ,
\end{equation}
and with $\pi_{\fA_1}: (\fA_1\cup \fA_2)^*\to \fA_1^*$ and $\pi_{\fA_2}: (\fA_1\cup \fA_2)^*\to \fA_2^*$ the projection maps.
The category $\cR\cL$ is the full subcategory of $\cF\cL$ where the objects are regular languages.
\end{defn}

As shown in \cite{FerMar} this defines an associative composition of morphisms. 

\smallskip

Since the main construction we are interested in considering in this paper, 
Boolean 1D TQFTs with defects associated to regular language as in \cite{GuImKaKhoLi},
uses a finite state automaton computing the language, rather than directly the language itself
(hence it depends on a choice, since different automata can compute the same language),
it is convenient here to reformulate this categorical construction in terms of automata.

\smallskip

At the level of automata, the notion corresponding to rational transductions is {\em transducers},
namely sets $\cT=(\fA,\fB, \fQ, F, \eta, q_0)$ consisting of an input alphabet $\fA$, an output alphabet $\fB$,
a set of states $\fQ$ with an initial state $q_0$ and a set $F$ of final states and a set of transitions
$\eta \subset \fQ \times \fA^* \times \fB^* \times \fQ$. Every rational transduction can be
computed by a transducer (Theorem~6.1 of \cite{Berstel}).

\smallskip

The rational transduction computed by a transducer $\cT=(\fB,\fC, \fQ, F, \eta, q_0)$ is given by the two
projections $\alpha: \fA^* \to \fB^*$ and $\beta: \fA^* \to \fC^*$, with $\cL_{reg}$ the regular language
of paths where one considers the set  $\cP(q,q')$ of paths $(q_1,b_1,c_1,q_2)\cdots (q_n,b_n,c_n,q_{n+1})$ with $q=q_1$,
$q'=q_{n+1}$ and $(q_i,b_i,c_i,q_{i+1})\in \eta$, and let $\cP(q_0,F)=\cup_{q\in F} \cP(q_0,q)$. 
The sets $\cP(q,q')$ are regular languages and so is $\cP(q_0,F)=\cL_{reg}$.

\smallskip

Conversely, given a rational transduction $X(\fA,\alpha,\beta,\cL_{reg})$ with $\alpha: \fA^* \to \fB^*$ and
$\beta: \fA^* \to \fC^*$, let $\cM=(\fQ,\fA,q_0,F,\tau)$ be a FSA that computes the regular language $\cL_{reg}$.
Then a transducer that computes $X$ is obtained by taking $\cT=(\fB,\fC,\fQ, F, \eta, q_0)$ with
$\eta=\{ (q, \alpha(a),\beta(a), q') \,|\, a\in \fA, \, (q,a,q')\in \tau \}$. 

\smallskip

We can view a transducer as a correspondence between automata in the following way.

\begin{prop}\label{TMprop}
The action of rational transductions as morphisms of context-free languages induces an action of transducers
on PSA and FSA automata, of the following form. 
Let $\cM$ be an automaton that is either a PSA $\cM=(\fQ_\cM, \fA_\cM, \Gamma_\cM, \tau_\cM, q_{0,\cM}, F_\cM, z_{0,\cM})$ 
or a FSA if $\cM=(\fQ_\cM, \fB_\cM, \tau_\cM, q_{0,\cM}, F_\cM)$. A transducer $\cT=(\fA_\cM,\fB, \fQ, F, \eta, q_0)$ 
with $\alpha: \fA^* \to \fA_\cM^*$ and
$\beta: \fA^* \to \fB^*$ maps $\cM$
to a new automaton (a PSA or a FSA, respectively) of the form
\begin{equation}\label{TMautom}
\begin{array}{l}
\cT(\cM) =(\fQ_{\cT(\cM)}, \fA_{\cT(\cM)}, \Gamma_{\cT(\cM)}, \tau_{\cT(\cM)}, q_{0,\cT(\cM)}, F_{\cT(\cM)}, 
z_{0,\cT(\cM)}) \\
\cT(\cM) =(\fQ_{\cT(\cM)}, \fA_{\cT(\cM)}, \tau_{\cT(\cM)}, q_{0,\cT(\cM)}, F_{\cT(\cM)}) \, ,
\end{array}
\end{equation}
where $\fA_{\cT(\cM)}=\fB$, $\Gamma_{\cT(\cM)}=\Gamma_\cM$, $z_{0,\cT(\cM)}=z_{0,\cM}$
and with set of states $(\fQ\times \fQ_\cM)\sqcup \fQ'$, where $\# \fQ' =\sum_{a\in \fA} n_a \, | \beta(a) |$,
with $|\beta(a)|$ the number of letters in $\fB$ of the word $\beta(a)$, and $n_a$ the number of pairs of
transitions in $\eta\times \alpha^{-1}(\tau_\cM)$ labelled by $a$.  The transitions in $\tau_{\cT(\cM)}$ are
subdivisions of paths $(q,\beta(a),z,q',w)$ with $q,q'\in \fQ\times \fQ_\cM$ into $| \beta(a) |$ steps with
endpoints in $\fQ'$. 
\end{prop}

\proof The construction of the automaton $\cT(\cM)$ follows the action of a corresponding 
rational transduction $X=X(\fA,\alpha,\beta,\cL_{reg})$, 
mapping a context-free language $\cL$ to a context free language $\beta(\alpha^{-1}(\cL)\cap \cL_{reg})$.
Suppose that $\cM$ is an automaton computing $\cL$. Then $\alpha^{-1}(\cL)$ is computed by
an automaton $\alpha^{-1}(\cM)$ which has the same data $\fQ_\cM$, $q_{0,\cM}$, $F_\cM$,
and the same $\Gamma_\cM$ and $z_{0,\cM})$ in the PSA case. The set $\tau_{\alpha^{-1}(\cM)}$ contains 
a transition $(q,a,q')\in \fQ_\cM\times \fA \times \fQ_\cM$ (in the FSA case) or $(q,a,z,q',w)\in \fQ_\cM\times \fA \times
\Gamma_\cM\times \fQ_\cM\times \Gamma_\cM^*$ (in the PSA case) iff the corresponding
$(q,\alpha(a),q')$ (respectively, $(q,\alpha(a),z,q',w)$) is a composition of transitions in $\tau_\cM$. 
The regular language $\cL_{reg}$ of the rational transduction is computed, as we recalled above,
as the regular language of paths of the transducer $\cT$. We denote by $\cM_\cT$ the FSA obtained
from the data of $\cT$, that computes $\cL_{reg}$. Then the language $\alpha^{-1}(\cL)\cap \cL_{reg}$
is computed by the product automaton 
$$ \cM_\cT \times \alpha^{-1}(\cM) =( \fQ \times \fQ_\cM, \fA, \Gamma_\cM, \tau_\cT \times \tau_{\alpha^{-1}(\cM)}, 
(q_{0,\cT},q_{0,\cM}), F_\cT \times F_\cM, z_0)\, . $$
Note that in the case of non-deterministic PSAs recognition by final state or by empty stack is equivalent,
so here above we consider the final states case. (The product automaton is in general nondeterministic.) 
This product can be seen as taking $\#\fQ$ copies of the PSA (or FSA) automaton $\alpha^{-1}(\cM)$
and adding the transitions coming from the FSA $\cM_\cT$ with states $\fQ$, between the corresponding
states in $\fQ \times \fQ_\cM$. Thus, a transition in $\cM_\cT \times \alpha^{-1}(\cM)$ is a pair of
transitions, one in $\cM_\cT$ and one in $\alpha^{-1}(\cM)$, which means that 
the words recognized by the resulting automaton are exactly those that are recognized by both $\alpha^{-1}(\cM)$ 
and $\cM_\cT$, hence those in the intersection $\alpha^{-1}(\cL)\cap \cL_{reg}$. Then, the image
$\beta(\alpha^{-1}(\cL)\cap \cL_{reg})$ is computed by an automaton $\beta(\cM_\cT \times \alpha^{-1}(\cM))$
that is obtained as follows. First consider the same automaton $\cM'=\cM_\cT \times \alpha^{-1}(\cM)$
and relabel transitions $(q,a,q')$ of $\cM'$ (or $(q,a,z,q',w)$ in the PSA case) as
$(q,\beta(a),q')$ (respectively, $(q,\beta(a),z,q',w)$). Then write out $\beta(a)=b_1\ldots b_k$ as
a word in the alphabet $\fB$. Add intermediate vertices (new states) along the edge corresponding to the
transition so that it becomes a composition of transitions $(q,b_1,q_1)\ldots (q_{k-1},b_k,q')$ in the FSA
case and $(q,b_1,z,q_1,z)(q_1,b_2,z,q_2,z)\ldots (q_{k-1},b_k, z, q', w)$ in the PSA case. 
We then define the image $\cT(\cM):=\beta(\cM_\cT \times \alpha^{-1}(\cM))$.
\endproof

\begin{prop}\label{AutoCat}
There is a category $\cA$ whose objects are automata $\cM$ (FSA and PSA as in Definition~\ref{FSAPSA})
and morphisms are correspondences given by transducers $\cT$, acting on automata as in Proposition~\ref{TMprop}.
For $\cT_i=(\fB_i,\fC_i,\fQ_i, F_i,\eta_i,q_{0,i})$, $i=1,2$, with homomorphisms 
$\alpha_i: \fA_i^* \to \fB_i^*$ and $\beta_i: \fA_i^* \to \fC_i^*$, and $\eta_i=\{(q,\alpha(a),\beta(a),q')\}$,
the composition $\cT_2\circ \cT_1$ is given by 
\begin{equation}\label{MTcomposeAct}
 (\cT_2\circ \cT_1)(\cM)= \beta( \cM_{\cT_2\circ \cT_1} \times \alpha^{-1}(\cM) )\, ,
\end{equation}
with $\alpha: (\fA_1\cup \fA_2)^* \to \fB_1^*$ and $\beta: (\fA_1\cup \fA_2)^* \to \fC_2^*$, with
$\alpha=\alpha_1 \circ \pi_{\fA_1}$ and $\beta =\beta_2\circ \pi_{\fA_2}$, and 
\begin{equation}\label{MTcomposeAut}
 \cM_{\cT_2\circ \cT_1} = \alpha_2^{-1} \beta_1 (\cM_{\cT_1}) \times \cM_{\cT_2} \, . 
\end{equation}
Let $\cF\cA$ denote the full subcategory of $\cA$ where the objects are finite state automata.
There is a functor from the category $\cA$ to the category $\cF\cL$ that assigns to an
automaton the language computed by the automaton and to a transducer the rational
transduction that it computes. It maps the subcategory $\cR\cA$ to the subcategory $\cR\cL$.
\end{prop}

\proof The composition of two transducer morphisms is given by $\cT_2\circ \cT_1$ mapping
an automaton $\cM$ to the automaton $(\cT_2\circ \cT_1)(\cM)$. 
To see that composition is well defined, we check that $\cT_2\circ \cT_1(\cM)$ defined as above 
agrees with 
$$ \cT_2(\cT_1(\cM))=  \beta_2( \cM_{\cT_2}   \times \alpha_2^{-1} (\beta_1(\cM_{\cT_1} \times \alpha_1^{-1}(\cM))) )\, . $$
Since $\beta=\beta_2 \circ \pi_{\fA_2}$, we have
$$ \beta_2(\cM_{\cT_2}\times \alpha_2^{-1}\beta_1(\cM_{\cT_1}))=\beta(\cM_{\cT_2\circ\cT_1}). $$
Moreover, we have $\beta_2 \alpha_2^{-1} \beta_1 \alpha_1^{-1}(\cM) =\beta \alpha^{-1}(\cM)$. In fact,
both automata have the same set of states $\fQ_\cM\sqcup \fQ'$ with
$\# \fQ' =\sum_{a\in \fA_1} n_a \, | \beta_1(a) |$,
with $|\beta_1(a)|$ the number of letters in $\fB_1$ of the word $\beta_1(a)$, and $n_a$ the number of pairs of
transitions in $\alpha^{-1}(\tau_\cM)$ labelled by $a$ (i.e., transitions in $\cM$ labelled by $\alpha(a)$). 
Moreover, for $\tilde w\in \fA_2^*$, the triple $(q,\beta_2(\tilde w),q')$ is a sequence of transitions in 
$\beta_2 \alpha_2^{-1} \beta_1 \alpha_1^{-1}(\cM)$ iff $(q,\alpha_2(\tilde w), q')=(q,\beta_1(w), q')$, with $w\in \fA_1^*$, 
is a sequence of transitions in $\beta_1 \alpha_1^{-1}(\cM)$, iff $(q,\alpha_1(w),q')$ is a sequence 
of transitions in $\cM$. Similarly, 
for $(w, \tilde w)\in (\pi_{\fA_1}, \pi_{\fA_2})((\fA_1\sqcup \fA_2)^*)=\fA_1^*\times\fA_2^*$
a triple $(q,\beta(w,\tilde w),q')=(q,\beta_2(\tilde w),q')$ is a sequence of transitions in $\beta \alpha^{-1}(\cM)$
iff $(q,\alpha(w,\tilde w),q')=(q,\alpha_1(w),q')$ is a sequence of transitions in $\cM$.
The associativity of the composition of transducers obtained in this way follows by a similar,
argument, as in the composition of rational transductions, with the same argument given in \cite{FerMar}.
Since the composition of transducers is designed to correspond to the composition of
rational transductions in the category $\cF\cL$, it 
follows that mapping an automaton $\cM\mapsto \cL_\cM$ to the language it computes and
mapping a transducer $\cT \mapsto X_\cT$ to the rational transduction it computes, defines a functor
$\Phi: \cA \to \cF\cL$. Regular languages are closed under intersection and image and preimage 
under homomorphisms
hence $\cT$ maps FSAs to FSAs, hence the subcategory $\cF\cA$ is mapped to $\cR\cL$.
\endproof

\smallskip

\section{Boolean 1D TQFTs with defects}\label{1DTQFTdefSec}

We summarize here briefly the construction of \cite{GuImKaKhoLi}, which associates to a
finite state automaton computing a regular language a Boolean 1D TQFT with defects.
We then analyze some aspects of this model that are not explicitly discussed in \cite{GuImKaKhoLi}, 
which include the structure of morphisms, the dependence on the automaton (rather than the language),
and the role of transductions. We follow here the approach to correspondences via transductions
developed in \cite{Berstel}, and the resulting categorical approach formulated in \cite{FerMar}, reviewed
in \S \ref{CatFLsec}.

\subsection{Finite State Automata and Boolean 1D TQFTs with Defects}

Recall that in general a TQFT is a functor from a category of cobordisms to a category of modules $R$-Mod,
for a commutative ring $R$ (or more generally with values in a symmetric monoidal category).

We recall the categories of one-dimensional cobordisms considered in \cite{GuImKaKhoLi}. 
In the 1D case, ${\rm Cob}$ denotes the category of one-dimensional oriented cobordisms, with objects 
given by $0$-manifolds, namely finite disjoint unions of two generating objects $\{ + , - \}$ (indicating a point with a positive or 
negative orientation), including the empty $0$-manifold $\emptyset$, 
and with morphisms that are 1-dimensional cobordisms, generated by the cup 
and cap cobordisms and the permutation cobordisms.  Functors $F: {\rm Cob} \to R$-Mod are
determined by finitely generated projective $R$-modules $M$, by setting $F(+)=M$, $F(-)=M^\vee=\Hom(M,R)$
with the cup and cap cobordisms mapped to the homomorphisms
$$ \cup: R \to M\otimes M^\vee \, , \ \ \ \ \cup(1)=\sum_i m_i \otimes m_i^\vee $$
for $\{ m_i \}$ a basis of $M$ (image of the standard basis of $R^n$ under the projection $p: R^n \to M$),
and $\{ m_i^\vee \}$ the dual basis, and with $\cap$ given by the evaluation morphism 
$\cap : M\otimes M^\vee \to R$. These satisfy the composition relations of the cap and cup cobordisms. 

\begin{figure}
    \centering
    \includegraphics[width=0.7\linewidth]{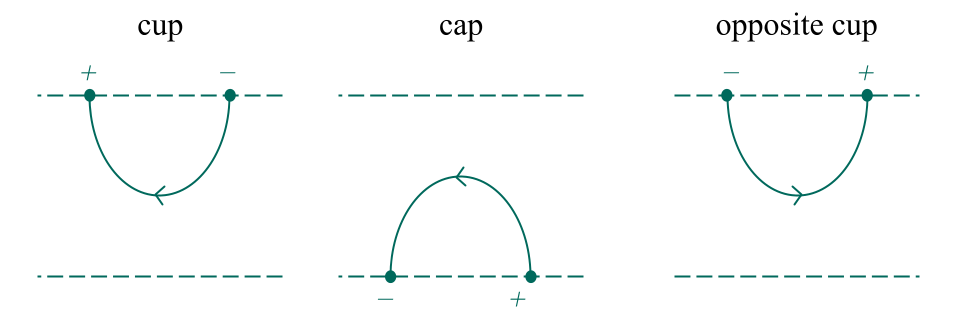}
    \caption{The generating morphisms of the category of one-dimensional cobordisms. The cup and cap morphisms and permutations which compose with them to get the opposite cup and cap, respectively.
    \label{morph}}
\end{figure}

The category ${\rm Cob}$ of 1D cobordisms is extended to a category that includes cobordisms with half-intervals,
or equivalently with ``floating endpoints". The resulting category ${\rm Cob}_1$ has the same objects as ${\rm Cob}$
and additional generating morphisms given by the half-intervals, seen as cobordisms between a point 
and the empty 0-manifold. Functors $F: {\rm Cob}_1 \to R$-Mod also map these half-edges cobordisms 
to either a morphism $m: M \to P$ (a choice of an element of $P$) or a morphism $m^\vee: M \to R$ 
(a choice of an element in $M^\vee$). 

The category ${\rm Cob}$ can also be extended to 1D cobordisms with defects, namely marked points
inserted in the one-dimensional cobordisms and marked by the elements of a finite set $\fA$. The
resulting category ${\rm Cob}_\fA$ has the same objects as ${\rm Cob}$, given by the $0$-manifolds, 
and morphisms the 1-dimensional cobordisms with finitely many marked points labelled by $\fA$. The
functors $F: {\rm Cob}_\fA \to R$-Mod map a line with a single marked point with label $a\in \fA$
to either an emdomorphism $T_a : M \to M$ or $T_a^\vee : M^\vee \to M^\vee$ depending on the
orientation. 

Finally, these two extensions of ${\rm Cob}$ can be combined into a category ${\rm Cob}_{\fA,1}$
that contains both floating endpoints and defects. 

\begin{figure}
    \centering
    \includegraphics[width=0.8\linewidth]{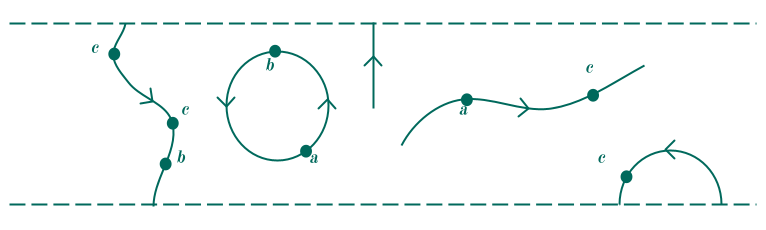}
    \caption{All four types of morphisms in the category of 1-dimensional cobordisms with defects, $Cob_{1,\sum}$, with $\sum=\{a,b,c\}$.
    \label{defmorph}}
\end{figure}

In \cite{GuImKaKhoLi}, one then considers, instead of TQFTs given by functors to a category $R$-Mod
of modules over a ring, a more general version where the target category is the category $\bB$-Mod of
seminodules over the Boolean semiring.

Then it is shown in \cite{GuImKaKhoLi} that a finite state automaton $\cM=(\fQ,\fA,\tau,q_0, F)$
determines a functor $F_\cM : {\rm Cob}_{\fA,1} \to \bB$-Mod, namely a Boolean 1D TQFT with defects.
The functor $F_\cM$ is acts by $F_\cM(+)=\bB\fQ$ (the free Boolean semimodule generated by the
set of states $\fQ$ of the automaton) and $F_\cM(+)=\bB\fQ^\vee$ with $\fQ^\vee=\{ q^\vee \}_{q\in \fQ}$ 
with $g^\vee(q')=\delta_{q,q'}$ the dual basis. Note here that we are implicitly assuming that the
set of states $\cM$ of the automaton is in a sense minimal for the transitions $(q,a,q')\in \tau$, 
namely $\fQ$ is the set of endpoints of transitions in $\tau$ and not larger. This will be relevant
in Lemma~\ref{PhialphaM} below where the set of states needs to be projected down to a minimal one. 
An object $\epsilon_1 \sqcup \cdots \sqcup \epsilon_n$
with $\epsilon_i=\pm$ is mapped to $F_\cM(\epsilon_1)\otimes \cdots \otimes F_\cM(\epsilon_n)$
with the $\cup$ and $\cap$ homomorphisms as above, while a line wth a defect $a\in \fA$ is
mapped to the endomorphism $T_a: \bB\fQ \to \bB\fQ$ defined by the transition functions $(q,a,q')\in \tau$
of the automaton $\cM$ by
$$ T_a : q \mapsto \sum_{(q,a,q')\in \tau} q' \, , $$
and its dual $T_a^\vee$ for the case with the opposite orientation.

\subsection{Transducers and functorial properties}

\begin{figure}
    \centering
    \includegraphics[width=0.8\linewidth]{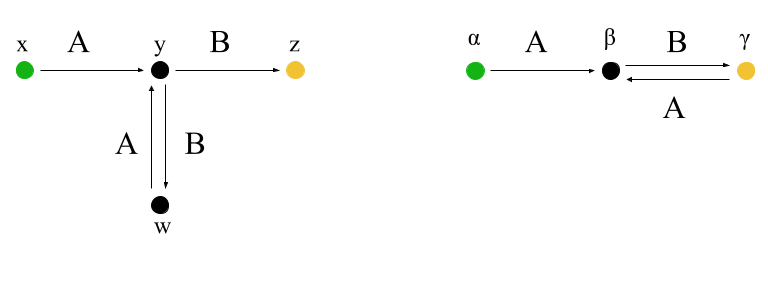}
    \caption{Two finite state automata which compute the same regular language $(AB)^n$.
    \label{AB}}
\end{figure}

The TQFT construction in \cite{GuImKaKhoLi}  is based on the automaton $\cM$ not on the regular
language it computes, in the sense that different automata computing the same language would
not necessarily determine the same TQFT.  It is then natural to ask if the information embedded in 
these TQFT's is in some way fundamental to the language itself. 
With two automata that compute the same language, it would seem natural to expect that their 
corresponding TQFTs should be naturally isomorphic. Simple examples, however, show machines 
that compute the same language but have a different number of states, hence quickly rule out
this possibility (Figure~\ref{AB}). A better approach is to obtain a functorial dependence of
the TQFT construction on the automaton, in such a way that morphisms of automata (given
by transducers) result in natural transformations of the associated TQFTs seen as functors
from the category ${\rm Cob}_{1,\fA}$ of 1D cobordisms with defects and the category
of $\bB$-semimodules over the Boolean semiring $\bB$. In order to obtain morphisms
of $\bB$-semimodules from transducers, we need to extend the category of $\bB$-semimodules
to morphisms given by correspondences, namely the category of spans ${\rm Span}(\bB\text{-Mod})$
of spans of the category $\bB\text{-Mod}$. 

\smallskip

We discuss here the functoriality properties of the 
assignment of a 1D TQFT with defects to a finite state automaton, as constructed in \cite{GuImKaKhoLi} 
and recalled above, when one considers FSAs as objects in the category $\cF\cA$ introduced above,
with morphisms given by transducers. Since TQFTs themselves are functors, it is natural to expect
that a good construction of TQFTs must itself satisfy functorial properties. 
The fact that morphisms in the category $\cF\cA$ are given by correspondences suggests that
a modification, of a similar nature, of the target category of the functors defining Boolean 1D TQFTs with defects.

\smallskip

Given a category $\cC$ with pullbacks, the associated category of spans ${\rm Span}(\cC)$ has
the same objects as $\cC$ and morphisms given by isomorphism classes of spans, namely, 
for $X,Y \in {\rm Obj}(\cC)$ a morphism in ${\rm Mor}_{{\rm Span}(\cC)}(X,Y)$ is an isomorphism class of 
a triple $(Z,f_X,f_Y)$ consisting of an object 
$Z\in {\rm Obj}(\cC)$ and two morphisms in $\cC$ forming a diagram
$$ \xymatrix{ & Z \ar[dl]^{f_X} \ar[dr]_{f_Y} & \\ X & & Y \, .} $$ 
Spans include ordinary morphisms in $\cC$, when $Z=X$.
The isomorphism classes are take with respect to 
morphisms of spans given by morphisms $Z\to Z'$ that form a commutative
diagram relating the two spans. The composition of spans is given by the pullback,
$$ \xymatrix{ & & Z\times_Y Z' \ar[dl]^{\pi_Z} \ar[dr]_{\pi_{Z'}} & &  \\
& Z \ar[dl]^{f_X} \ar[dr]_{f_Y} & & Z'  \ar[dl]^{f_Y} \ar[dr]_{f_W}  & \\ X & & Y  & & W } $$ 
and taking isomorphism classes is necessary since the pullback is defined only up to 
canonical isomorphism. 

\smallskip

The category of semimodules over a semiring has pullbacks \cite{Taka}, hence we can consider
the category ${\rm Span}(\bB\text{-Mod})$ that has as objects the semimodules over
the Boolean semiring $\bB$ and spans as morphisms. Thus, we assume here the following
variant of the definition of Boolean 1D TQFT with defects of \cite{GuImKaKhoLi}.

\begin{defn}\label{SpanTQFT}
A Boolean 1D TQFT with defects with alphabet $\cA$ is a functor 
$\Phi: {\rm Cob}_{1,\fA} \to {\rm Span}(\bB\text{-{\rm Mod}})$.
A {\rm strict} Boolean 1D TQFT with defects with alphabet $\cA$ is a functor 
$\Phi: {\rm Cob}_{1,\fA} \to \bB\text{-{\rm Mod}}$.
\end{defn}

Note that the Boolean 1D TQFTs with defects considered in \cite{GuImKaKhoLi} also satisfy
our definition, since ordinary morphisms in $\bB\text{-Mod}$ can also be seen as spans. 
The only difference is that we only consider them up to isomorphisms. 

\begin{lem}\label{alphaCob}
A monoid homomorphism $\alpha: \fA^* \to \fB^*$ induces a functor $\varphi_\alpha: {\rm Cob}_{1,\fA} \to {\rm Cob}_{1,\fB}$
and a functor of 1D TQFTs with defects, 
\begin{equation}\label{alphaTQFTs}
\alpha^*: {\rm Func}({\rm Cob}_{1,\fB},{\rm Span}(\bB\text{-{\rm Mod}})) \to {\rm Func}({\rm Cob}_{1,\fA},{\rm Span}(\bB\text{-{\rm Mod}}) )
\end{equation}
\end{lem}

\proof The functor is the identity on objects and on the cup and cap cobordisms.  
It sends a line with a marked
point with label $a\in \fA$ to a line with $k$ marked points (along the orientation of the line) with labels
$b_1, \ldots, b_k \in \fB$ for $\alpha(a)=b_1 \cdots b_k \in \fB^*$. This assignment sends a composition
of cobordisms with defects to the composition of their images. 
Given a monoid homomorphism $\alpha: \fA^* \to \fB^*$ the associated functor 
$\varphi_\alpha: {\rm Cob}_{1,\fA} \to {\rm Cob}_{1,\fB}$ then induces by precomposition a functor
$\alpha^*$ as in \eqref{alphaTQFTs}, with $\alpha^*(\Phi)=\Phi\circ\varphi_\alpha$.
\endproof

\smallskip

\begin{lem}\label{PhialphaM}
Let $\cM$ be a FSA and let $\Phi_\cM\in {\rm Func}({\rm Cob}_{1,\fB}, \bB\text{-{\rm Mod}})$ denote the
1D TQFT with defects constructed in \cite{GuImKaKhoLi} (recalled in \S \ref{1DTQFTdefSec} above).
For a monoid homomorphism $\alpha: \fA^* \to \fB^*$, the functor $\Phi_{\alpha^{-1}(\cM)}
\in {\rm Func}({\rm Cob}_{1,\fA}, \bB\text{-{\rm Mod}})$ is given by $\Phi_{\alpha^{-1}(\cM)}=\Pi_\alpha \, \alpha^*(\Phi_\cM)$,
where $\Pi_\alpha: \bB\fQ_\cM \to \bB\fQ_{\cM,\alpha}$ is the projection onto the submodule 
spanned by the subset $\fQ_{\cM,\alpha}\subset \fQ_\cM$ of states of the automaton $\cM$ that are endpoints
of sequences of transitions of the form $(q,\alpha(a),q')$, for $a\in \fA$. The functor $\Phi_{\alpha^{-1}(\cM)}$
acts on objects by $\Phi_{\alpha^{-1}(\cM)}(\underline{\epsilon})=\bB\fQ_{\cM,\alpha}^{\epsilon_1}\otimes \cdots \otimes \bB\fQ_{\cM,\alpha}^{\epsilon_m}$, for an object $\underline{\epsilon}=(\epsilon_1,\ldots,\epsilon_m)$, $\epsilon_i\in \{ \pm \}$, and
on cobordisms $\fc \in {\rm Mor}_{{\rm Cob}_{1,\fA}}(\underline{\epsilon},\underline{\epsilon'})$ as
$\Pi_\alpha \, \Phi(\varphi_\alpha(\fc))\, \Pi_\alpha$. 
\end{lem}

\proof Transitions in $\alpha^{-1}(\cM)$ are of the form $(q,a,q')$ where $(q,\alpha(a),q')$ is a sequence of
transitions in $\cM$. Thus, the set of states of $\alpha^{-1}(\cM)$ that are involved in transitions in the
automaton is given by the subset $\fQ_{\cM,\alpha}\subset \fQ_\cM$. 
Thus, the functor $\Phi_{\alpha^{-1}(\cM)}$ acts on objects by 
$$ \Phi_{\alpha^{-1}(\cM)}(\underline{\epsilon})=
\bB\fQ_{\cM,\alpha}^{\epsilon_1}\otimes \cdots \otimes \bB\fQ_{\cM,\alpha}^{\epsilon_m} =\Pi_\alpha \alpha^*\Phi(\underline{\epsilon}) \, . $$
For the action on cobordisms, it suffices to check the generators, namely the cap and cup cobordisms,
the half-lines, and the lines with defects. For the cap and cup cobordisms we have
$\Phi_{\alpha^{-1}(\cM)}(\cup)(q^*\otimes q')=\Phi(\cup)(\Pi_\alpha(q)^* \otimes \Pi_\alpha(q'))
=\Pi_\alpha(q)^*(\Pi_\alpha(q')))= \delta_{q,q'}$ for $q,q'\in \fQ_{\cM,\alpha}$ and zero otherwise,
and $\Phi_{\alpha^{-1}(\cM)}(\cap)(1)=\Pi_\alpha \sum_{q\in \fQ_\cM} q^* \otimes q 
= \sum_{q\in \fQ_{\cM,\alpha}} q^* \otimes q$. The half-lines correspond to either $1 \mapsto q_0$
or $q \mapsto q_F^*(q)$ (assuming the automata have a single final state $q_F$), where both
$q_0, q_F \in \fQ_{\cM,\alpha}$. The lines with defects will only contain 
defects of the form $\alpha(a)=b_1 \ldots b_n$ hence to transitions $(q, \alpha(a), q')$ in $\cM$
between $q,q'\in \fQ_{\cM,\alpha}$, and one such line $\fc$ with defect $\alpha(a)$
maps to $\Phi_{\alpha^{-1}(\cM)}(\fc)(q)=\sum_{q'\in \fQ_{\cM,\alpha},\, (q,a,q')\in \tau_{\alpha^{-1}(\cM)}} q' =
\Pi_\alpha\Phi_{\alpha^{-1}(\cM)}(\fc)\Pi_\alpha(q)$. 
\endproof

\smallskip

To take into account the observation of Lemma~\ref{PhialphaM}, we will modify our
definition of the pullback in the following way.

\smallskip

\begin{defn}\label{pullback2}
For a functor $\Phi : {\rm Cob}_{1,\fB} \to \bB\text{-{\rm Mod}}$ and a monoid homomorphism $\alpha: \fA^* \to \fB^*$,
Let $\Phi(+)_\alpha$ denote the largest subspace of $\Phi(+)\in \bB\text{-{\rm Mod}}$ that is invariant under all the
$\Phi(\fc_{\alpha(a)})$, for $a\in \fA$, where $\fc_w$ denotes the line with defects $w=b_1\ldots b_k$ in $\fB^*$. 
We then define the modified pullback, which we still denote for simplicity $\alpha^*\Phi$ as the functor
$\alpha^*\Phi: {\rm Cob}_{1,\fA} \to \bB\text{-{\rm Mod}}$ that maps objects by $\alpha^*\Phi(+)=\Phi(+)_\alpha$
and $\alpha^*\Phi(-)=(\Phi(+)_\alpha)^\vee$, with $\alpha^*\Phi(\underline{\epsilon})=\Phi(\underline{\epsilon})_\alpha$ 
defined accordingly, and that maps morphisms as $\alpha^*\Phi(\fc)=\Pi_\alpha \Phi(\fc\circ \alpha) \Pi_\alpha$,
indicating the restriction of $\Phi(\fc\circ \alpha)$ to the subspaces $\Phi(\underline{\epsilon})_\alpha$. 
\end{defn}

\smallskip

\begin{thm}\label{transdNT}
For a FSA $\cM$ let $\Phi_\cM\in {\rm Func}({\rm Cob}_{1,\fB}, \bB\text{-{\rm Mod}})$ denote the
1D TQFT with defects constructed in \cite{GuImKaKhoLi} (and recalled in \S \ref{1DTQFTdefSec} above).
A transducer $\cT=(\fB,\fC,\fQ,F,\eta,q_0)$ with $\alpha: \fA^* \to \fB^*$ and $\beta: \fA^*\to \fC^*$
determines a natural transformation $\gamma_\cT: \alpha^* \Phi_{\cM} \to \beta^* \Phi_{\cT(\cM)}$
between the Boolean 1D TQFT with defects  $\alpha^*\Phi_{\cM}, \beta^*\Phi_{\cT(\cM)}\in 
{\rm Func}({\rm Cob}_{1,\fA},{\rm Span}(\bB\text{-{\rm Mod}}))$, with the modified pullbacks as in Definition~\ref{pullback2}.
\end{thm}

\proof The functor $\Phi_\cM\in {\rm Func}({\rm Cob}_{1,\fB}, \bB\text{-{\rm Mod}})$ constructed in \cite{GuImKaKhoLi}
determines as a functor in ${\rm Func}({\rm Cob}_{1,\fB},{\rm Span}(\bB\text{-{\rm Mod}}))$, by viewing the morphisms in
$\bB\text{-{\rm Mod}}$ as spans, and passing to isomorphism classes. As in Definition~\ref{pullback2} and 
Lemma~\ref{PhialphaM}, we have $\alpha^* \Phi_{\cM}(\underline{\epsilon})=\bB\fQ_{\cM,\alpha}^{\epsilon_1}\otimes \cdots 
\otimes \bB\fQ_{\cM,\alpha}^{\epsilon_m}=\Phi_{\alpha^{-1}(\cM)}$, if we use the modified pullback of 
Definition~\ref{pullback2}, and similarly for $\beta^* \Phi_{\cT(\cM)}$ and $\Phi_{\beta^{-1}(\cT(\cM))}$. 
For a given object $\underline{\epsilon}$, we take 
$\gamma_{\cT,\underline{\epsilon}}: \Phi_{\alpha^{-1}(\cM)}(\underline{\epsilon})\to 
\Phi_{\beta^{-1}(\cT(\cM))}(\underline{\epsilon})$
to be the span 
\begin{equation}\label{spanT}
\xymatrix{  
& \bB(\fQ_\cT\times \fQ_{\cM})^{\epsilon_1}\otimes \cdots \otimes \bB(\fQ_\cT\times \fQ_{\cM})^{\epsilon_m} \ar[dl]^{f_{\cM, \alpha, \underline{\epsilon}}} \ar[dr]_{f_{\cT(\cM)}, \beta, \underline{\epsilon}} & \\
\bB\fQ_{\cM}^{\epsilon_1}\otimes \cdots \otimes \bB\fQ_{\cM}^{\epsilon_m} & & \bB\fQ_{\cT(\cM)}^{\epsilon_1}\otimes \cdots \otimes \bB\fQ_{\cT(\cM)}^{\epsilon_m} \, . } 
\end{equation}
The morphism is of the form $f_{\cM,\alpha,\underline{\epsilon}}=\otimes_i f_{\cM,\alpha}^{\epsilon_i}$ where
by $f_{\cM,\alpha}^{\pm}$ acts a $\Pi_\alpha$ to project $\bB(\fQ_\cT\times \fQ_{\cM})^{\pm}$ onto the subspace $\bB(\fQ_\cT\times \fQ_{\cM,\alpha})^{\pm}$ and then maps $\bB(\fQ_\cT\times \fQ_{\cM,\alpha})^{\pm}$ 
to the subspace $\bB\fQ_{\cM,\alpha}^\pm$ inside $\bB\fQ_{\cM}^{\pm}$. The set of states $\fQ_{\cT(\cM)}$ is given by 
$\fQ_{\cT(\cM)}=(\fQ_\cT\times \fQ_\cM)\sqcup \fQ'$ with  $\# \fQ' =\sum_{a\in \fA} n_a \, | \beta(a) |$,
with $|\beta(a)|$ the number of letters in $\fB$ of the word $\beta(a)$, and $n_a$ the number of pairs of
transitions in $\eta\times \alpha^{-1}(\tau_\cM)$ labelled by $a\in \fA$.  Note, however, that because
the coefficients are in the Boolean semiring, these integer multiplicities do not appear and the Boolean 
sum of the coefficients is equal to $1$. The morphism is of the form
$f_{\cT(\cM),\beta,\underline{\epsilon}}=\otimes_i f_{\cT(\cM),\beta}^{\epsilon_i}$ where
$f_{\cT(\cM),\beta}^{\pm}$ acts as $\Pi_{\beta}$ to project onto $\bB((\fQ_\cT\times \fQ_\cM)_\beta)^{\pm}$ and then 
maps this to the subspace $\bB\fQ_{\cT(\cM)_\beta}^\pm$ 
inside $\bB((\fQ_\cT\times \fQ_\cM)\sqcup \fQ')^{\pm}=\bB(\fQ_{\cT(\cM)})^\pm$. 
To check that this defines a natural transformation we need to check the commutativity of
the diagram of morphisms in the category ${\rm Span}(\bB\text{-{\rm Mod}})$:  
$$ \xymatrix{  
& \bB(\fQ_\cT\times \fQ_{\cM})^{\epsilon_1}\otimes \cdots \otimes \bB(\fQ_\cT\times \fQ_{\cM})^{\epsilon_m}
\ar[dd]^{\Phi_T(\fc)}
 \ar[dl]^{f_{\cM, \alpha, \underline{\epsilon}}} \ar[dr]_{f_{\cT(\cM), \beta, \underline{\epsilon}}} & \\
\bB\fQ_\cM^{\epsilon_1}\otimes \cdots \otimes \bB\fQ_\cM^{\epsilon_m} \ar[dd]^{\alpha^* \Phi_{\cM}(\fc)} & & \bB\fQ_{\cT(\cM)}^{\epsilon_1}\otimes \cdots \otimes \bB\fQ_{\cT(\cM)}^{\epsilon_m} \ar[dd]^{\beta^*\Phi_{\cT(\cM)}(\fc)} \\
& \bB(\fQ_\cT\times \fQ_\cM)^{\epsilon'_1}\otimes \cdots \otimes \bB(\fQ_\cT\times \fQ_\cM)^{\epsilon'_{m'}} \ar[dl]^{f_{\cM, \alpha, \underline{\epsilon'}}} \ar[dr]_{f_{\cT(\cM), \beta, \underline{\epsilon'}}} & \\
\bB\fQ_\cM^{\epsilon'_1}\otimes \cdots \otimes \bB\fQ_\cM^{\epsilon'_{m'}} & & \bB\fQ_{\cT(\cM)}^{\epsilon'_1}\otimes \cdots \otimes \bB\fQ_{\cT(\cM)}^{\epsilon'_{m'}} \, , } 
$$
for $\fc$ a morphism given by a 1-dimensional cobordism in ${\rm Cob}_{1,\fA}$ between the 
objects $\underline{\epsilon}$ and $\underline{\epsilon'}$. It suffices to check this on the 
generators of morphisms, namely the cup and cap cobordism, the half-lines, and the
lines with defects. The arrow $\Phi_T(\fc)$ in the middle of the diagram is defined in the usual way
for cup, cap, half-line cobordisms, and as a sum over transitions
$((q_T, \alpha(a),\beta(a),q_T'),(q,\alpha(a),q'))$ in $\fQ_T\times \fQ_\cM$. 

By construction of the modified pullback, $\alpha^*\Phi_\cM(\fc_a)$ maps
the range of $f_{\cM,\alpha,\underline{\epsilon}}$ to the range of $f_{\cM,\alpha,\underline{\epsilon'}}$. 
Indeed, for $\fc_a$ a line with a single defect marked by $a$, 
in the automaton $\alpha^{-1}(\cM)$ the generators of this type of cobordisms are lines marked by $\alpha(a)$, for $a\in \fA$,
so that $\fc_a$ corresponds to transitions $(q,\alpha(a),q')$ in $\fQ_{\cM,\alpha}$. Similarly for $\beta^*\Phi_{\cT(\cM)}(\fc)$
mapping the range of $f_{\cT(\cM), \beta, \underline{\epsilon}}$
to the range of $f_{\cT(\cM), \beta, \underline{\epsilon'}}$, with 
transitions in $\cT(\cM)$ labelled by $\beta(a)$ that map $\fQ_{\cT(\cM)_\beta}$ to itself.
The left-side of the diagram commutes
since projecting the image of $\Phi_T(\fc_a) (q,q_T)=\sum_{((q_T, \alpha(a),\beta(a),q_T'),(q,\alpha(a),q'))} (q',q'_T)$
to $\fQ_{\cM,\alpha}$ is (up to a multiplicity that does not appear in the Boolean coefficients) the same as
the image $\alpha^*\Phi_\cM(\fc_a)(q)=\sum_{(q,\alpha(a),q'))} q'$ with $q\in \fQ_{\cM,\alpha}$. 
Similarly, the projection to $\fQ_{\cT(\cM),\beta}$ is (up to a multiplicity not visible with Boolean coefficients) 
the same as the image $\beta^*\Phi_{\cT(\cM)}(q,q_T)=\sum_{(q_T, \beta(a),q_T')} (q',q'_T)$ where
$(q,q_T)\in \fQ_T\times \fQ_\cM=\fQ_{\cT(\cM),\beta}$. 

In the case of the cup and cap cobordisms, $\alpha^*\Phi_\cM(\cup)$ and $\alpha^*\Phi_\cM(\cap)$ act as 
the restrictions of $\Phi_\cM(\cup)$ and $\Phi_\cM(\cap)$ to 
the range of $f_{\cM,\alpha,\underline{\epsilon}}$. These agree with the projections 
of $\Phi_T(\cup)$ and $\Phi_T(\cap)$, and similarly for the restriction of $\beta^*\Phi_{\cT(\cM)}$ to
the range of  $f_{\cT(\cM), \beta, \underline{\epsilon}}$. 
The half-lines map $1 \mapsto q_0$ and $q\mapsto q_F^*(q)$ (assuming the automata have a unique final state $q_F$) with
$q_0,q_F \in \fQ_{\cM,\alpha}$ lifting to the initial and final state for $\fQ_T$ and mapping to
initial and final state for $\fQ_{\cT(\cM)}$. 
\endproof

\smallskip

\subsection{Subregular classes}

In this section we consider in particular certain classes of subregular languages
that are of interest in the theory of formal languages, and that we discuss here
from the point of view of the properties of the corresponding 1D Boolean TQFTs. 
A discussion of classes of subregular languages is given in \cite{JagRog}.
Subregular languages are often characterized in model-theoretic terms 
by logical definability in terms of certain classes of logical expressions, or
computationally by tests on the strings of the language, usually describable in
terms of $k$-factors and $k$-subsequences. We show here that for these
subregular classes of languages their computational description determines
additional structure on the corresponding TQFTs. 

\smallskip

A $k$-factor is a set of $k$ adjacent symbols in strings $w=a_1,\ldots, a_N \in \cL$, with 
added symbols $\rtimes, \ltimes$ to mark beginning and ending of strings. For example,
the language $(AB)^n$ computed by the automata in Figure~\ref{AB} is characterized by
the $2$-factors $\{ \rtimes A, AB, BA, B\ltimes \}$, meaning that only these 2-factors can
occur as two consecutive letters in any word $w=ABABAB \ldots$ in the language. 

\smallskip

Languages that can be characterized in this way are called {\em strictly local} (SL), and if
characterized in terms of a list of $k$-factors they are in the class SL$_k$. 

\smallskip

In terms of the corresponding TQFT, consider first the case of SL$_2$ languages like
the $(AB)^n$ language above. The fact that, out of all the possible $2$-factors $aa'$ in the
alphabet $\fA$ only a certain subset $\cF_2(\cL)$ is admissible in the words of $\cL$ means that, in
the construction of the TQFT, any composition of two lines with defects resulting in a
marking of defects by two consecutive letters $aa' \notin\cF_2(\cL)$ results in operators
$T_a, T_{a'}$ with the property that $T_a T_{a'}=0$. For example, in the case of the $(AB)^n$
language, the only $2$-factors (not involving $\rtimes,\ltimes$) that are not admissible are $BB$ and $AA$, 
and this means that $T_B$ is a coboundary operator ($T_B^2=0$) on any of the spaces 
$\Phi_{\cM}(\underline{\epsilon})=\bB\fQ_{\cM}^{\epsilon_1}\otimes \cdots \otimes \bB\fQ_{\cM}^{\epsilon_m}$,
and so is $T_A$ (which also satisfies $T_A^2=0$). 
This means that the associated TQFT has an additional structure in this case, which
is a cohomological structure given either by $H_B(\Phi_{\cM}(\underline{\epsilon}),T_B):={\rm Ker}(T_B)/{\rm Image}(T_B)$
or by $H_A(\Phi_{\cM}(\underline{\epsilon}),T_A):={\rm Ker}(T_A)/{\rm Image}(T_A)$. One can combine both
as the space $H_B(\Phi_{\cM}(\underline{\epsilon}),T_B)\oplus H_A(\Phi_{\cM}(\underline{\epsilon}),T_A)$. 
This is true more generally of any language $\cL$ in SL$_2$, with the resulting cohomological structures
$$ \oplus_{ab\notin \cF_2(\cL)} H_{ab}(\Phi_{\cM}(\underline{\epsilon}),T):= {\rm Ker}(T_a)/ {\rm Image}(T_b)\, . $$
Note that if we also want to include the $2$-factors with $\rtimes,\ltimes$ we would have to also exclude
$\rtimes B$ and $A \ltimes$. However, from the TQFT point of view is is more natural to only consider
$k$-factors without $\rtimes,\ltimes$, with only the $2$-factors $\cF_2((AB)^n)=\{ AB, BA \}$.

\smallskip

In the case of SL$_k$ languages, with $k\geq 2$, where testing $w\in \cL$ is determined by the
list of admissible $k$-factors $w=a_1\ldots a_k \in \cF_k(\cL)$, one can similarly think of the
complementary set of the non-admissible $k$-factors $w=a_1\ldots a_k \notin \cF_k(\cL)$ as
defining a family of cohomology structures 
$$ \oplus_{w \notin \cF_2(\cL)} \oplus_{w=w_1 w_2}  H_{w_1,w_2}(\Phi_{\cM}(\underline{\epsilon}),T):= 
{\rm Ker}(T_{w_1})/ {\rm Image}(T_{w_2})\, , $$
where $T_w=T_{a_1}\cdots T_{a_k}$, and where $T_w=\Phi_{\cM}(\fc)$ with
$\fc=\fc_1 \circ \cdots \circ \fc_k$ and $\fc_i$ the cobordism given by a line wth defect marked by the letter $a_i$. 

\smallskip

Another class of subregular languages is the {\em locally testable} (LT) languages, where in addition
to admissible $k$-factors, one also requires the actual occurrence of certain $k$-factors in the words $w\in \cL$.
Thus, the characterization of $w\in \cL$ in an LT$_k$ language is in terms of a collection of Boolean conditions of the form
$(\neg w_1) \land \cdots \land (\neg w_n) \land w_1' \land \cdots \land w_m'$ for a collection $\{ w_i,w_j' \}_{n,m}$ of $k$-factors,
with the $w_i$ being the inadmissible $k$-factors and the $w_j'$ being the $k$-factor that are required to occur. 
This means that, in addition to the cohomological conditions coming from the inadmissible $k$-factors, as in the SL-case,
one also has an additional structure for LT$_k$ languages. Namely, let $\cR_k(\cL)=\{ w'_j \}_{j=1\ldots,m}$ be
the collection of the required $k$-factors that characterize $\cL$ (we can ignore the inadmissible $k$-factors
as those we already know from the SL-case give rise to cohomology). 
Given objects $\underline{\epsilon}$ and $\underline{\epsilon'}$ consider the set $\cC:={\rm Cob}_{\cR_k(\cL)}(\underline{\epsilon},\underline{\epsilon'})$ of cobordisms $\fc\in {\rm Mor}_{{\rm Cob}_{1,\fA}}(\underline{\epsilon},\underline{\epsilon'})$ between them consisting of lines with defects marked by words $w \in \fA^*$ that contain the required
$k$-factors in $\cR_k(\cL)$. This set of words has the form 
$w\in \cup_r \fA^* u_1 \fA^* u_2 \fA^* \cdots u_r \fA^*$, with $u_i\in \cR_k(\cL)$.
This set of cobordisms $\cC$ has two communting actions of the endomorphism monoids $\cE(\underline{\epsilon})$ and $\cE(\underline{\epsilon'})$ of the two objects $\underline{\epsilon},\underline{\epsilon'}$, 
$$ \cE(\underline{\epsilon})={\rm Cob}_{1,\fA}(\underline{\epsilon},\underline{\epsilon}) \ \ \ \text{ and } \ \ \ 
\cE(\underline{\epsilon'})={\rm Cob}_{1,\fA}(\underline{\epsilon'},\underline{\epsilon'}) \, , $$
respectively acting by precomposition and postcomposition. 

\smallskip

Unlike the case of the SL languages where the characterization in terms of admissible $k$-factors is
local, which translates in terms of cobordisms into a condition on the compositions of defects,
in the case of LT languages one sees that there is no possibility of characterizing the language in
terms of local conditions on composition of cobordisms that correspond to successive defects. 
The only way to describe the LT conditions is in terms of a global object, namely the set $\cC$
described above that includes cobordisms with arbitrarily large number of defects. The language
is recovered by those  floating lines (compositions
of two half-lines with endpoints inside the cobordism, see Figure~\ref{defmorph} inside the 
cobordisms in the set $\cC$, for which $q_F^*(T_w(q_0))=1 \in \bB$, namely $w\in \cL$. 

\smallskip

There are several interesting subclasses of regular languages 
(see, for instance, \cite{JagRog} and \cite{Rawski})
that have a natural characterization in
terms of logic, in addition to the Strictly Local (SL) and 
Locally Testable (LT), discussed briefly here. These include
classes known as
Non-Counting (NC), Locally Threshold Testable (LTT), Piecewise Testable
(PT), and Strictly Piecewise (SP). It is natural to ask how the logic
properties that single out these subclasses of regular languages translate into
properties of the associated TQFT  and whether they admit a
possible path integral interpretation. We will not discuss this
question further in the present paper, and we leave it to future work.

\smallskip

\section{1D TQFTs: from Regular to Context-free Languages}

\subsection{Categorical finite state automata}

In \S \ref{catFLsec} we have recalled the construction of \cite{FerMar} of a category of
formal languages and a corresponding category of automata. In fact, the automata themselves
can be generalized to a categorical formulation, as in \cite{MelZei}. We discuss here how
to extend the construction of \cite{FerMar} that we just recalled, in this more general categorical
framework.

\smallskip

This classical definition of FSAs can be generalized in a categorical framework as
in \cite{MelZei}, in the following form. (We assume, for simplicity, that FSAs have a unique final
state, $F=q_f$, since one can always modify the automaton to have this property without altering
the language computed.)

\begin{itemize}
\item Recall that a functor $f: \cC_1 \to \cC_2$ between two categories is {\em finitary} if the preimage
of every object and every morphism is finite: $\# f^{-1}(Y)<\infty$ for each $Y\in {\rm Obj}(\cC_2)$
and $\# f^{-1}(\varphi)<\infty$ for each $\varphi \in {\rm Mor}_{\cC_2}$.
\item A functor $f: \cC_1 \to \cC_2$ has the {\em unique lifting of factorizations} (ULF) property if 
whenever $f(\psi)=\varphi_2 \circ \varphi_1$ in
${\rm Mor}_{\cC_2}$, there are unique $\psi_1,\psi_2 \in {\rm Mor}_{\cC_1}$ with $\psi=\psi_2\circ\psi_1$
and $f(\psi_i)=\varphi_i$. 
\end{itemize}

\begin{defn}\label{FSAPSAcat}
A {\rm categorical} finite state automaton is a datum $\cM=(\cQ,\cC,\tau: \cQ\to \cC,q_0, q_f)$,
where $\cQ$ and $\cC$ are categories and $\tau: \cQ\to \cC$ is a functor that is finitary and has
the ULF property. Objects of $\cQ$ are the states of the automaton and morphisms of $\cQ$ are
the {\rm runs} (concatenations of transitions) of the automaton. The two objects $q_0, q_f$ are the
initial and final state, and the {\em regular language of arrows} $\cL_\cM$ computed by the
automaton $\cM$ is the set 
$$ \cL_\cM =\{ \varphi\in {\rm Mor}_\cC \,|\, \varphi = \tau(w) \text{ for some } w\in {\rm Mor}_\cQ(q_0.q_f) \}\, ,$$
that is, morphisms $\varphi \in {\rm Mor}_\cC$ that lift to runs of the automaton from inital to final state.
\end{defn}

\begin{rem}\label{regregcat}{\rm
A language  $\cL \subset \fA^*$ is regular language in the usual sense if it is a regular language $\cL=\cL_\cM$,
in the sense of Definition~\ref{FSAPSAcat}, for $\cQ$ the category with ${\rm Obj}(\cQ)=\fQ$
and $\cC=\cB_\fA$ the category with a single object and ${\rm Mor}_{\cB_\fA}=\fA^*$, with $\tau$
mapping transitions $(q,a,q')$ of $\fQ$ arrows $a\in \fA$ in $\cB_\fA$. }
\end{rem}

\smallskip

We can then extend the construction of the associated Boolean 1D TQFT with defects to
this categorical version of finite state automata, in the following way. Given a small category $\cQ$,
let $\bB{\rm Obj}(\cQ)$ denote the set of finitely supported
functions $f: {\rm Obj}(\cQ) \to \bB$, which we can write as sums $f=\sum_q a_q \, \delta_q$,
where all but finitely many coefficients $a_q\in \bB$ vanish. 

\smallskip

\begin{thm}\label{catM1DTQFT}
Let $\cM=(\cQ,\cC,\tau: \cQ\to \cC,q_0, q_f)$ be a categorical finite state automaton as in
Definition~\ref{FSAPSAcat}. The automaton $\cM$ determines a Boolean 1D TQFT with
defects $\Phi_\cM: {\rm Cob}_{1,{\rm Mor}_\cC} \to \bB$-Mod, where 
\begin{itemize}
\item {\em Objects}: $\Phi_\cM(+)=\bB{\rm Obj}(\cQ)$
and $\Phi_\cM(+)=\bB{\rm Obj}(\cQ)^\vee$, and $\underline{\epsilon}=(\epsilon_i)_{i=1}^n$ with
$\epsilon_i\in \{ \pm \}$ is mapped to $\Phi_\cM(\underline{\epsilon})=\bB{\rm Obj}(\cQ)^{\epsilon_1}\otimes \cdots \otimes 
\bB{\rm Obj}(\cQ)^{\epsilon_n}$ with $\bB{\rm Obj}(\cQ)^+=\bB{\rm Obj}(\cQ)$ and $\bB{\rm Obj}(\cQ)^-=\bB{\rm Obj}(\cQ)^\vee$.
\item {\em Morphisms}:  The image of a line with a single defect
labelled by an element $\varphi\in {\rm Mor}_\cC(x,y)$, for two $x,y\in {\rm Obj}(\cC)$ is given by 
$\Phi_\cM(\varphi)=T_\varphi: \bB{\rm Obj}(\cQ) \to \bB{\rm Obj}(\cQ)$
\begin{equation}\label{Tphi}
T_\varphi:  \sum_q a_q \delta_q \mapsto \sum_{q,q'} \sum_{w\in {\rm Mor}_\cQ(q,q')} \, a_q \, 
 \delta_{\tau(w),\varphi} \, q'  \, . 
\end{equation}
\item The images under $\Phi_\cM$ of the $\cup$ and $\cap$ cobordisms and of the half-lines are
as in the case of the ordinary finite state automata.
\end{itemize}
\end{thm}

\proof The argument is the same as in the original construction of  \cite{GuImKaKhoLi}, except for
the change to the structure of defects and composition of lines with defects. Given $\varphi_1,\varphi_2 \in {\rm Mor}_{\cC}$
a composed cobordism given by a line with two successive defects labelled by $\varphi_1$ and $\varphi_2$
gives rise to the composition of operators $T_{\varphi_2}\circ T_{\varphi_1}$. Note that, by \eqref{Tphi}, this
composition is identically zero if $\varphi_1\in {\rm Mor}_{\cC}(x,y)$ and $\varphi_1\in {\rm Mor}_{\cC}(y',z)$
where $y\neq y'$. Indeed, if that is the case, then $\tau^{-1}(y)\cap \tau^{-1}(y')=\emptyset$ in ${\rm Obj}(\cQ)$
and the condition $\tau(w)=\varphi_1$ in $T_{\varphi_1}$ implies $\tau(q)=x$ and $\tau(q')=y$, while applying
$T_{\varphi_2}$ sets $\tau(q')=y'$. Thus, $T_{\varphi_2}\circ T_{\varphi_1}$ can be nontrivial only when $y=y'$,
that is, when $\varphi_2\circ \varphi_1$ is defined. Then \eqref{Tphi} implies that 
$$ T_{\varphi_2}\circ T_{\varphi_1}( \sum_q a_q \delta_q) = \sum_{q,q''} \sum_{w_1 \in {\rm Mor}_\cQ(q,q')} \sum_{w_2\in {\rm Mor}_\cQ(q',q'')} 
\, a_q \, \delta_{\tau(q), x} \,  \delta_{\tau(w_1),\varphi_1} \delta_{\tau(w_2),\varphi_2} \, q^{\prime\prime} $$
$$ = \sum_{q,q''} \sum_{w\in {\rm Mor}_\cQ(q',q'')} \, a_q \, \delta_{\tau(q), x} \,  \delta_{\tau(w),\varphi_2\circ \varphi_1}  \, q^{\prime\prime}
=T_{\varphi_2\circ \varphi_1}( \sum_q a_q \delta_q) \, . $$
Note that we are using both the finitary condition on the functor $\tau$, for the sums in \eqref{Tphi}, and the
ULF condition, for the identification 
$\delta_{\tau(w_1),\varphi_1} \delta_{\tau(w_2),\varphi_2} =  \delta_{\tau(w),\varphi_2\circ \varphi_1}$ in the composition.
\endproof

\smallskip

The correspondences between finite state automata given by transducers can also be generalized to
this categorical formulation of finite state automata in the form of pullback constructions. 

\begin{defn}\label{CatTransd}
A categorical transducer is a datum $\cT=(\cQ_T,\cC,\cC',\eta=(\alpha,\beta): \cQ_T\to \cC \times \cC')$,
where the functor $\alpha: \cQ_T\to \cC$ is injective-on-objects, and also $\alpha$ and $\beta$ are
finitary and with the ULF condition.
A transducer acts on a categorical finite state automaton $\cM=(\cQ_\cM,\cC,\tau: \cQ_\cM\to \cC,q_0, q_f)$ by 
\begin{equation}\label{catTM}
 \cT(\cM)=(\cQ_\cM \times_{\cC,\tau,\alpha} \cQ_T, \pi^*(\beta): \cQ_\cM \times_{\cC,\tau,\alpha} \cQ_T \to \cC', \beta(\alpha^{-1}(\tau(q_0))), \beta(\alpha^{-1}(\tau(q_f)))) \, , 
\end{equation} 
where $\cQ_\cM \times_{\cC,\tau,\alpha} \cQ_T$ is the pullback with 
$$ {\rm Obj}(\cQ_\cM \times_{\cC,\tau,\alpha} \cQ_T)=\{ (q,q_T)\in {\rm Obj}(\cQ)\times {\rm Obj}(\cQ_T)
\,|\, \tau(q)=\alpha(q_T) \} $$
$$ {\rm Mor}_{\cQ_\cM \times_{\cC,\tau,\alpha} \cQ_T}((q,q_T),(q',q'_T))=\{ (w,\hat w)\in {\rm Mor}_{\cQ}(q,q')
\times {\rm Mor}_{\cQ_T}(q_T,q'_T) 
\,|\, \tau(w)=\alpha(\hat w) \} \, . $$
\end{defn}

\smallskip

We can view a categorical transducer acting on a categorical FSA as a diagram of categories and functors
$$ \xymatrix{ \cQ\times_{\cC,\tau,\alpha} \cQ_T \ar[r]^{\quad\pi} \ar[d]  & \cQ_T \ar[r]^\beta \ar[d]^\alpha & \cC' \\
\cQ \ar[r]^\tau & \cC & } $$
where the condition that $\alpha$ is injective-on-objects preserves the property that automata have a single
initial and final state, 
$$ {\rm Obj}(\cQ\times_{\cC,\tau,\alpha} \cQ_T)=\{ (q,q_T)\,|\, \tau(q)=\alpha(q_T) \}=\{ (q, \alpha^{-1}(\tau(q)) ) \, |\, q\in {\rm Obj}(\cQ) \} \cong {\rm Obj}(\cQ)\, . $$
On morphisms, correspondingly, we have
$$ {\rm Mor}_{\cQ\times_{\cC,\tau,\alpha} \cQ_T}(q,q')=\{ (w,\hat w)\in {\rm Mor}_\cQ(q,q')\times {\rm Mor}_{\cQ_T}(\alpha^{-1}(\tau(q)), \alpha^{-1}(\tau(q'))) \,|\, \tau(w)=\alpha(\hat w) \} \, , $$
which we also write with the notation
$$ {\rm Mor}_{\cQ\times_{\cC,\tau,\alpha} \cQ_T}(q,q')= {\rm Mor}_\cQ(q,q') \times_{\tau,\alpha} {\rm Mor}_{\cQ_T}(\alpha^{-1}(\tau(q)), \alpha^{-1}(\tau(q'))) \, . $$

\smallskip

As in the case of ordinary FSAs with transducers as morphisms, we can form a category that
has categorical FSAs as objects and categorical transducers as morphisms.

\begin{thm}\label{CatCatMT}
The categorical finite state automata $\cM=(\cQ_\cM,\cC,\tau: \cQ_\cM\to \cC,q_0, q_f)$ with morphisms
given by categorical transducers $\cT=(\cQ_T,\cC,\cC',(\alpha,\beta))$
form a category.
\end{thm}

\proof The associativity of composition of morphisms is the main property to check. A composition of two
categorical transducers is obtained by considering a diagram
$$ \xymatrix{  &   \cQ_T \times_{\cC',\tau',\alpha'} \cQ_{T'} \ar[r]^{\quad\pi'} \ar[d]^{\pi_T} & 
\cQ_{T'} \ar[r]^{\beta'} \ar[d]^{\alpha'} & \cC'' \\
\cQ\times_{\cC,\tau,\alpha} \cQ_T \ar[r]^{\quad\pi} \ar[d]  & \cQ_T \ar[r]^\beta \ar[d]^\alpha & \cC' & \\
\cQ \ar[r]^\tau & \cC & & } $$
where the composition of categorical transducers is given by
$$ \cT'\circ \cT:=(  \cQ_T \times_{\cC',\tau',\alpha'} \cQ_{T'} , \cC, \cC'', (\alpha\circ \pi_T, \beta' \circ \pi')) \, . $$
To obtain the action of the transducer $\cT'\circ \cT$ on $\cM$ we 
complete the diagram by forming the pullback 
$$ \xymatrix{    \cQ\times_{\cC,\tau,\alpha\circ \pi_T} (\cQ_T \times_{\cC',\tau',\alpha'} \cQ_{T'}) \ar[d] \ar[r] & 
 \cQ_T \times_{\cC',\tau',\alpha'} \cQ_{T'}  \ar[d]^{\alpha \circ \pi_T} \ar[r]^{\quad\quad \beta'\circ \pi'} & \cC'' \\
\cQ \ar[r]^\tau & \cC  & } $$
This has set of objects 
$$ {\rm Obj}(  \cQ\times_{\cC,\tau,\alpha\circ \pi_T} (\cQ_T \times_{\cC',\tau',\alpha'} \cQ_{T'}) )
=\{ (q,\alpha^{-1}\tau(q), {\alpha'}^{-1}\beta \alpha^{-1}\tau(q)) \,|\, q\in {\rm Obj}(\cQ) \} \cong {\rm Obj}(\cQ) $$
and morphisms
$$ {\rm Mor}_{ \cQ\times_{\cC,\tau,\alpha\circ \pi_T} (\cQ_T \times_{\cC',\tau',\alpha'} \cQ_{T'})}(q,q') \cong  $$ $$ 
{\rm Mor}_\cQ(q,q')\times_{\tau,\alpha} {\rm Mor}_{\cQ_T} ( \alpha^{-1}\tau(q), \alpha^{-1}\tau(q')) \times_{\beta, \alpha'}
{\rm Mor}_{\cQ_{T'}} ({\alpha'}^{-1}\beta \alpha^{-1}\tau(q), {\alpha'}^{-1}\beta \alpha^{-1}\tau(q')) =  $$ 
$$  \bigcup_{w\in \alpha^{-1}(u)\, ,\, u\in {\rm Mor}_\cQ(q,q')} \{ (w,\hat w) \,|\, \beta(w) =\alpha'(\hat w) \} = 
\bigcup_{u\in {\rm Mor}_\cQ(q,q')} \{ u \} \times  {\alpha'}^{-1}\beta \alpha^{-1}\tau(u)\, . $$

By the pasting law for pullbacks, in the diagram
$$ \xymatrix{  \cQ\times_{\cC,\tau,\alpha\circ \pi_T} (\cQ_T \times_{\cC',\tau',\alpha'} \cQ_{T'})
\ar[r] \ar[d] &   \cQ_T \times_{\cC',\tau',\alpha'} \cQ_{T'} \ar[r]^{\quad\pi'} \ar[d]^{\pi_T} & 
\cQ_{T'}  \ar[d]^{\alpha'}  \\ 
\cQ\times_{\cC,\tau,\alpha} \cQ_T \ar[r]^{\quad\pi}  & \cQ_T \ar[r]^\beta  & \cC' } $$
the right square is a pullback, which implies that the combined square is a pullback iff the left one is.
The description above of objects and morphisms shows that we can also identify
$ \cQ\times_{\cC,\tau,\alpha\circ \pi_T} (\cQ_T \times_{\cC',\tau',\alpha'} \cQ_{T'})$ with 
$(\cQ\times_{\cC,\tau,\alpha} \cQ_T) \times_{\cQ_T,\pi,\pi_T} (\cQ_T \times_{\cC',\tau',\alpha'} \cQ_{T'})$
which is a pullback. Thus, $ \cQ\times_{\cC,\tau,\alpha\circ \pi_T} (\cQ_T \times_{\cC',\tau',\alpha'} \cQ_{T'})$ 
is also the pullback 
$(\cQ\times_{\cC,\tau,\alpha} \cQ_T)\times_{\cC', \beta\circ\pi, \alpha'} \cQ_{T'}$ that describes
the action of the transducer $\cT'=(\cQ_{T'}, \cC', \cC'', (\alpha',\beta'))$ on the image $\cT(\cM)$. 
Thus, we obtain that composition is well defined: $(\cT'\circ\cT)(\cM)=\cT'(\cT(\cM))$.  Consider then
three categorical transducers $\cT$, $\cT'$, and $\cT''$ forming a diagram
$$ \xymatrix{ & & \cQ_{T''} \ar[d]^{\alpha''} \ar[r]^{\beta''} & \cC''' \\
& \cQ_{T'} \ar[d]^{\alpha'} \ar[r]^{\beta'} & \cC'' & \\
\cQ_T \ar[d]^\alpha \ar[r]^\beta & \cC' & \\ \cC & & & } $$
To check associativity 
$(\cT'' \circ\cT')\circ \cT=\cT''\circ(\cT'\circ \cT)$
we need to match the actions of these categorical transducers on categorical FSAs. Both are
described by compositions of pullback diagrams, 
$$ \xymatrix{  \cQ_T\times_{\cC', \beta, \alpha'\circ \pi'} (\cQ_{T'}\times_{\cC'',\beta',\alpha''} \cQ_{T''}) \ar[r]  \ar[dd]
  & 
\cQ_{T'}\times_{\cC'',\beta',\alpha''} \cQ_{T''} \ar[r] \ar[d]^{\pi'} & \cQ_{T''} \ar[d]^{\alpha''} \ar[r]^{\beta''} 
& \cC''' \\
& \cQ_{T'} \ar[d]^{\alpha'} \ar[r]^{\beta'} & \cC'' & \\
\cQ_T \ar[d]^\alpha \ar[r]^\beta & \cC' & \\ \cC & & & }  $$
and
$$ \xymatrix{ ( \cQ_T\times_{\cC', \beta, \alpha'} \cQ_{T'}) \times_{\cC'', \beta'\circ \pi', \alpha''} \cQ_{T''} \ar[rr] \ar[d]
 & & \cQ_{T''} \ar[d]^{\alpha''} \ar[r]^{\beta''} & \cC''' \\
\cQ_T\times_{\cC', \beta, \alpha'} \cQ_{T'} \ar[r]^{\pi'} \ar[d] & \cQ_{T'} \ar[d]^{\alpha'} \ar[r]^{\beta'} & \cC'' & \\
\cQ_T \ar[d]^\alpha \ar[r]^\beta & \cC' & \\ \cC & & & }
$$
The comparison proceeds in this case as in the case of the compositions $\cT'(\cT(\cM))$ and $(\cT'\circ \cT)(\cM)$
discussed above, giving the corresponding identifications. 
\endproof

We can then extend the functorial mapping from the category of automata to the Boolean TQFTs
to the case of categorical automata and transducers. The categorical setting simplifies the 
result with respect to the case of classical automata discussed earlier in this paper. In particular,
unlike in Theorem~\ref{transdNT}, here, under the assumption that the categorical transducers have 
$\alpha: \cQ_T \to \cC$ injective-on-objects (which guarantees unique initial and final states
for the categorical automata), we do not need to use modified pullbacks and we do not
need to extend morphisms of $\bB$-semimodules passing to the category of spans. 

\begin{thm}\label{nattransCatMPhiM}
For a categorical FSA $\cM=(\cQ_\cM,\cC,\tau: \cQ_\cM\to \cC,q_0, q_f)$, let $$\Phi_\cM\in {\rm Func}({\rm Cob}_{1,{\rm Mor}_\cC}, \bB\text{-Mod})$$ be the associated 1D Boolean TQFT with defects as in Theorem~\ref{catM1DTQFT}. 
A categorical transducer $\cT=(\cQ_T,\cC,\cC', (\alpha,\beta))$, with $\alpha: \cQ_T \to \cC$ injective on objects,
determines a natural transformation
$\gamma_T: \alpha^* \Phi_\cM \to \beta^* \Phi_{\cT(\cM)}$. 
\end{thm}

\proof First observe that the functors $\alpha: \cQ_T \to \cC$ and $\beta: \cQ_T \to \cC'$ induce functors
(which we still denote in the same way) $\alpha: {\rm Cob}_{1,{\rm Mor}_{\cQ_T}} \to {\rm Cob}_{1,{\rm Mor}_{\cC}}$
and $\beta: {\rm Cob}_{1,{\rm Mor}_{\cQ_T}} \to {\rm Cob}_{1,{\rm Mor}_{\cC'}}$ that are the identity on objects
and on all generators of cobordisms except the line with defect $w\in {\rm Mor}_{\cQ_T}$ which is mapped to the
line with defect $\alpha(w)\in {\rm Mor}_{\cC}$, or $\beta(w)\in {\rm Mor}_{\cC'}$, respectively. Thus, we can
consider the pullbacks $\alpha^* \Phi_\cM$ and $\beta^* \Phi_{\cT(\cM)}$ in ${\rm Func}({\rm Cob}_{1,{\rm Mor}_{\cQ_T}}, \bB\text{-Mod})$. For an object $\underline{\epsilon}=(\epsilon_i)$ the images are given by
$$ \alpha^* \Phi_\cM(\underline{\epsilon})=\otimes_i \bB {\rm Obj}(\cQ_\cM)^{\epsilon_i} \ \ \text{ and } \ \ 
 \beta^* \Phi_{\cT(\cM)}(\epsilon)=\otimes_i \bB {\rm Obj}(\cT(\cQ_\cM))^{\epsilon_i} \, , $$
 with the notation as before, 
 with basis elements written in the form $\delta_q^+=\delta_q$ and $\delta_q^-=\delta_{q^\vee}$.
Since the functor $\alpha$ is injective-on-objects we have an identification $q \mapsto (q,\alpha^{-1}\tau(q))$,
$$ {\rm Obj}(\cQ_\cM) \stackrel{\cong}{\to} {\rm Obj}(\cQ_\cM \times_{\cC,\tau,\alpha} \cQ_T) =  {\rm Obj}(\cT(\cQ_\cM))\, . $$
We define $\gamma_{T,\underline{\epsilon}}: \alpha^* \Phi_\cM(\underline{\epsilon}) \to  \beta^* \Phi_{\cT(\cM)}(\epsilon)$
to simply map $\gamma_{T,\underline{\epsilon}}: \otimes_i \delta_{q_i}^{\epsilon_i} \mapsto \otimes_i \delta_{(q_i, \alpha^{-1}\tau(q_i))}^{\epsilon_i}$. To check that this defines a natural transformation, we only need to check the
case of cobordisms that are lines with defects, namely, for $w\in {\rm Mor}_{\cQ_T}(\hat q,\hat q')$, the commutativity
of the diagram 
$$ \xymatrix{ \bB {\rm Obj}(\cQ_\cM) \ar[d]^{T_{\alpha(w)}} \ar[r]^{\gamma_{T,+}\quad} & 
\bB{\rm Obj}(\cT(\cQ_\cM)) \ar[d]^{T_{\beta(w)}}\\
\bB {\rm Obj}(\cQ_\cM)  \ar[r]^{\gamma_{T,+}\quad } & 
\bB{\rm Obj}(\cT(\cQ_\cM)) } $$
On a basis element $\delta_q$ the morphism $T_{\alpha(w)}$ acts by
$$ T_{\alpha(w)}(\delta_q)=\sum_{q'} \sum_{u\in {\rm Mor}_{\cQ_\cM}(q,q')} \delta_{\tau(u),\alpha(w)} \, \delta_{q'} \, . $$
On the other hand, the morphism $T_{\beta(w)}$ acts on a basis element $\delta_{(q,\alpha^{-1}\tau(q))}$ as
$$ T_{\beta(w)}(\delta_{(q,\alpha^{-1}\tau(q))}) = \sum_{(q', \alpha^{-1}\tau(q'))} \,\,\,
\sum_{\hat u \in {\rm Mor}_{\cT(\cM)}((q,\alpha^{-1}\tau(q)), (q', \alpha^{-1}\tau(q'))} \delta_{\pi(\hat u),w}\,\, \delta_{(q', \alpha^{-1}\tau(q'))} \, , $$
with the projection $\pi: \cQ_\cM \times_{\cC,\tau,\alpha} \cQ_T \to \cQ_T$.  The set of morphisms of $\cT(\cM)$ in
the sum above is given by
 $$ \{ \hat u=(u,w)\in {\rm Mor}_\cQ(q,q')\times {\rm Mor}_{\cQ_T}(\alpha^{-1}(\tau(q)), \alpha^{-1}(\tau(q'))) \,|\, \tau(u)=\alpha(w) \} \, . $$
 Thus, the summation above is the same as the summation
 $$ \sum_{q'} \sum_{u \in {\rm Mor}_\cQ(q,q')} \delta_{\tau(u),\alpha(w)}\,\, \delta_{(q', \alpha^{-1}\tau(q'))} $$
 which shows that the diagram commutes.
 \endproof

\smallskip

This categorical version of finite state automata and regular languages was extended to
a categorical formulation of context-free grammars and context-free languages in 
\cite{MelZei}, \cite{MelZei2}, in the form of a categorical reformulation of the 
Chomsky--Sch\"utzenberger theorem representing context-free languages in terms of
regular languages as Dyck languages on matching sets of parentheses.

\subsection{Operads of spliced arrows and species}

We recall the following setting on species, colores operads, and 
operads of spliced words in a category from \cite{MelZei}, \cite{MelZei2}.

Let $C$ be a set (of colors) and let $V$ be a finite set of vertices
with assigned valencies (one can think of then as corollas consisting of
a vertex $v$ and $n=\deg(v)$ half-edges attached to it). The half-edges
are directed, with a single outgoing half-edge and all the other half-edges
oriented as incoming. One can assign colors to the half-edges by functions
$C \stackrel{o}{\leftarrow} V \stackrel{\iota}{\rightarrow} C^*$, 
where for a vertex with $\deg(v)=n$ one has $\iota : v \mapsto \iota(v)\in C^{n-1}$,
the vector of colors assigned to the incoming directions. The datum
$\bS=(C,V,\iota,o)$ is called a {\em species} (colored non-symmetric species). 
It is finite if both $C$ and $V$ are finite. A morphism for species is a commutative diagram
relating the respective $\iota,o$ maps. 

A free operad $\cO$ on a generating set of operations ${\rm Op}(\cO)$ is the
operad whose $n$-ary operations $\cO(n)$ are all the trees with $n$ leaves and
internal vertices of valence $k+1$ ($k$ inputs and one output) labelled by 
$k$-ary operations in ${\rm Op}(\cO)$.

Given a species  $\bS=(C,V,\iota,o)$ one obtains a colored operad $\cO_\bS$ 
which is the free operad with ${\rm Op}(\cO_\bS)(n)=\{ v \in V\,|\, \deg(v)=n+1 \}$ and 
with the partially defined operad compositions obtained by grafting output 
to input half-edges with matching colors. This assignment is functorial.
There is also a functor from operad to species that, to a given colored operad $\cO$ 
assigns a species $\bS_\cO$ with $V$ the set of operations in the operad with the
same colors. These are adjoint functors. 

Given a category $\cC$ one can form the {\em operad of spliced
arrows} $\cW\cC$ where the set of colors is the set of pairs $(X,Y)$ of
objects in $\cC$ and an operation in $\cW\cC(n)$ with inputs
colored by $(X_i, Y_i)$, $i=1,\ldots, n$ and output colored by $(X,Y)$ is a
sequence of the form
\begin{equation}\label{spliceword}
 w_0\, \square \, w_1 \, \square \, \cdots \, \square \, w_n 
\end{equation} 
with $w_i \in {\rm Mor}_\cC(Y_i, X_{i+1})$, for $i=0,\ldots, n$ with $Y_0=X$ and $X_{n+1}=Y$,
and where the $n$ $\square$ spaces in the sequence denote gaps, which are the inputs  
that can be filled by the operad composition. The composition operations splice another word
of this form into each of the square gaps of a given word, with the $(X,Y)$-colored identity
given by the word ${\rm id}_X \, \square \, {\rm id}_Y$. 

\begin{figure}
    \begin{center}
    \includegraphics[scale=0.6]{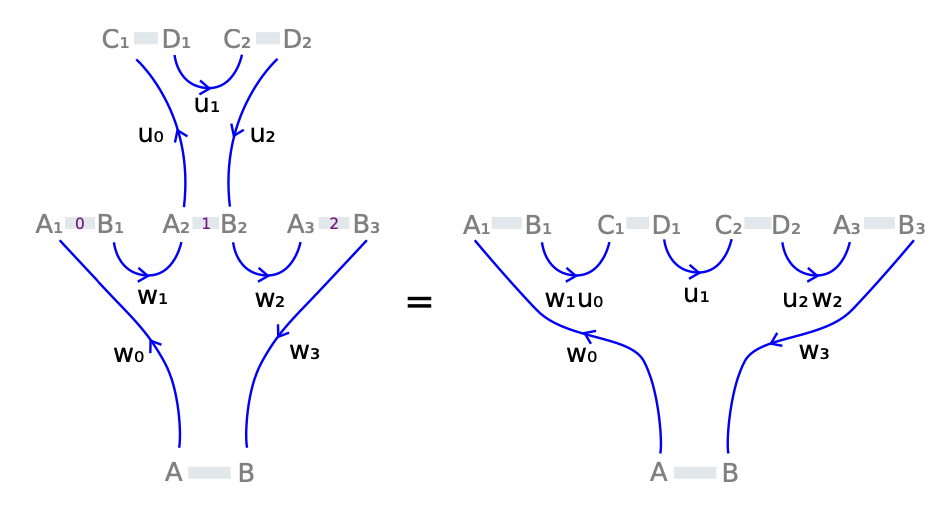} 
          \caption{Operad composition $f \circ_1 g$ of the sequence $g=u_0\square u_1 \square u_2$ into
       the sequence $f= w_0 \square w_1 \square w_2 \square w_3$, from \cite{MelZei2}.
    \label{SpliceOpFig}}
    \end{center}
\end{figure}

A context-free grammar $\cG=(\fN, \fA, P, S)$ has production rules $p \in P$ of the form
\begin{equation}\label{productionCF}
 p : N \mapsto w_0 N_1 w_1 \ldots N_n w_n 
\end{equation} 
where $N, N_i \in \fN$ are nonterminal symbols and $w_i \in \fA^*$ are words in the alphabet $\fA$
(terminal symbols). As observed in \cite{MelZei}, \cite{MelZei2}, a context-free grammar can therefore
be seen as a morphism of operads, 
\begin{equation}\label{operadFunc}
 P_\cG : \cO_\bS \to \cW\fA \, . 
\end{equation}  
Here $\cO_\bS$ is the free colored operad generated by the species $\bS$ determined by the set $P$ of production rules
with the set of colors given by the set $\fN$ of nonterminal symbols (including the start symbol $S$): for $p\in P$
as above, 
$$ \iota(p)=(N_1, \ldots, N_n) \ \ \ \text{ and } \ \ \  o(p)=N \, . $$
The operad $\cW\fA$ is the splice operad of the category $\cB_\fA$ that has a single object and 
morphisms generated by the letters $s\in \fA$, that is, ${\rm Mor}_{\cB_\fA}=\fA^*$; namely, the 
operad of words of the form \eqref{spliceword} in the alphabet $\fA$. In order to obtain ``constants"
in $\cW\fA$, namely filled in words in $\fA^*$ with no more spaces, we also need to include operations
in the species where the output is a terminal, rather than just a nonterminal.  

This rephrasing of a context-free grammar $\cG=(\fN, \fA, P, S)$ then makes it possible to
define categorical versions of context-free grammars, as in \cite{MelZei}, \cite{MelZei2}, given by
data $\cG^{cat}=( \cC, \bS, S, P)$ with $\cC$ a category, $\bS$ a finite species, and $P$ a morphism of operads
$P : \cO_\bS \to \cW\cC$. The language $\cL_\cC$ computed by a categorical context-free grammar
is the set of morphisms of the category $\cC$ that are in the image of $P$ (when seen as 
constants in $\cW\cC$, meaning with no remaining spaces $\square$). It is called the {\em language
of arrows} in $\cC$. 

Various properties of context-free grammars (linear, Chomsky normal form, bilinear, unambiguous) can
be phrased in terms of this categorical formulation (see \S 1.4 of \cite{MelZei2}). 

\subsection{Operad structure and cobordisms with defects}

In order to extend to context free grammars the categories of FSAs and transducers and their
functorial mapping to 1D TQFTs with defects, we first need to lift the mapping of a categorical
FSA automaton $\cM=(\cQ,\cC, \tau: \cQ\to \cC, (q_0,q_f))$ to the associated 1D TQFT
$\Phi_\cM:  {\rm Cob}_{1,{\rm Mor}_\cC} \to \bB\text{-{\rm Mod}}$  to the
level of operads.

\begin{figure}
    \begin{center}
    \includegraphics[scale=0.42]{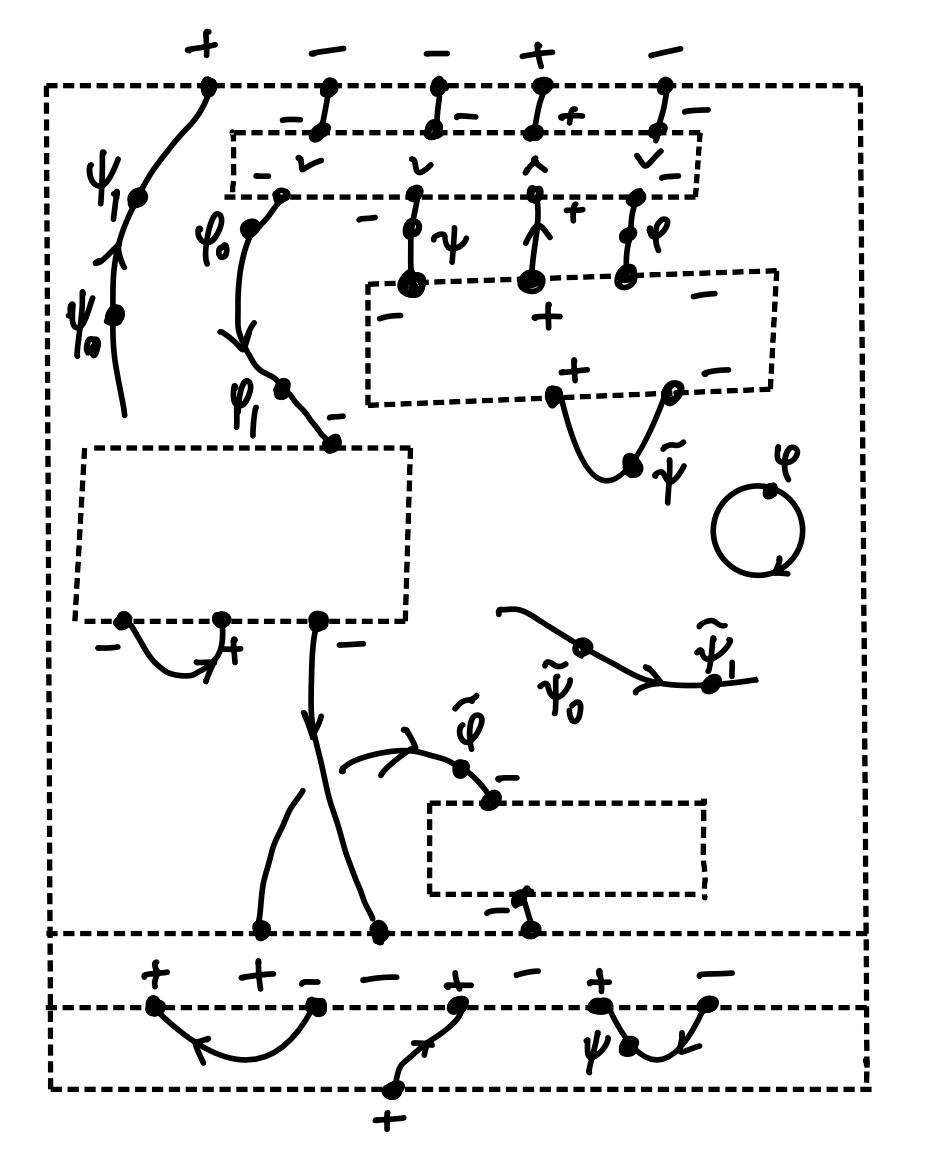}     
       \caption{Operations in the operad $\cO_{{\rm Cob},\cC}$ of 1D cobordisms with defects.
    \label{CobOperDefFig}}
    \end{center}
\end{figure}

\begin{defn}\label{OpCobCdef}
The colored operad $\cO_{{\rm Cob},\cC}$ of cobordisms with defects labelled by
morphisms of $\cC$ has $\cO_{{\rm Cob},\cC}(n)$ given by 1D cobordisms with defects
with $n$ square/rectangular boxes cut out (see Figure~\ref{CobOperDefFig}), with each box marked
by a color, given by a pair $(\underline{\epsilon},\underline{\epsilon'})$ of objects in the category
${\rm Cob}_{1,{\rm Mor}_\cC}$ respectively placed as in the figure. The operad composition consists
of plugging in other such cobordism wth matching boundary data into the boxes.
\end{defn} 

This operad structure of cobordisms with defects of Definition~\ref{OpCobCdef} is closely related 
to the operad of cobordisms considered, for instance, in \cite{Spivak}. 

\begin{defn}\label{operRMod}
Given a commutative ring $R$ and the category $R$-Mod of modules over $R$, the colored operad
$\cO_{R{\rm Mod}}$ has colors given by pairs of modules $(M,N)$ with $M=\otimes_{i\i \cI} M_i$ and $N=\otimes_{i\in \cI} N_i$
and the operations in $\cO_{R{\rm Mod}}(n)$ are of the form $\otimes_i \psi_{i,0}\square \cdots \square \psi_{i,n_i}$
(including the case of a single $\psi_{i,0}$ with no $\square$'s) with a total number of $n=\sum_i n_i$ boxes,
where $\psi_{i,j}\in {\rm Mor}_{R\text{-{\rm Mod}}}$. The operad composition fills in a box in a sequence
$\ldots \psi_{i,j}\square \psi_{i,j+1} \ldots$ with $\psi_{i,j}\in {\rm Mor}_{R\text{-{\rm Mod}}}(M_{i,j}, N_{i,j})$
and $\psi_{i,j+1}\in {\rm Mor}_{R\text{-{\rm Mod}}}(M_{i,j+1}, N_{i,j+1})$ where $N_{i,j}=\otimes_{a\in A} N_a$
and $M_{i,j+1}=\otimes_{a\in A} M_a$, with some  $\otimes_a \psi_{a,0}\square \cdots \square \psi_{a,n_a}$.
\end{defn}

One can consider the analogous constructions for semimodules over a semiring instead of modules over a ring.

\begin{defn}\label{splicedTQFT}
A {\em spliced TQFT} is a morphism of operads between the operads of spliced arrows
$$ \Psi: \cW {\rm Cob} \to \cW R\text{-{\rm Mod}} $$
of a cobordism category and a category of modules. A Boolean 1D spliced TQFT with defects in $\cC$
is a morphism of operads
$$ \Psi: \cW {\rm Cob}_{1,{\rm Mor}_\cC} \to \cW\bB\text{-{\rm Mod}}\, , $$
with defects in an auxiliary category $\cC$. 
An operadic Boolean 1D TQFT with defects in $\cC$ is a morphism of operads
$$ \Psi: \cO_{{\rm Cob},\cC} \to \cO_{\bB{\rm Mod}}\, . $$
\end{defn}

\begin{figure}
    \begin{center}
    \includegraphics[scale=0.42]{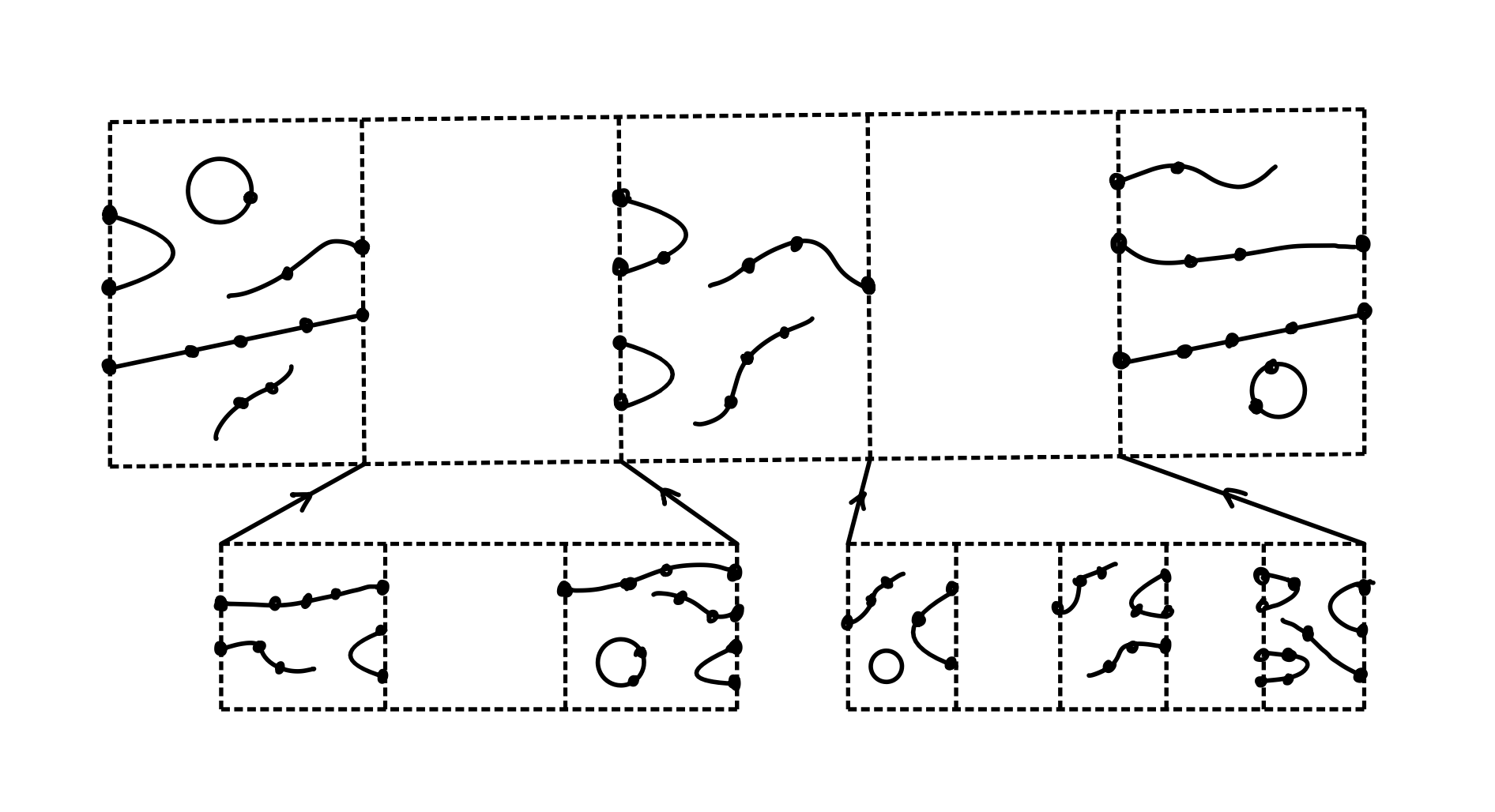}     
       \caption{1D cobordisms with defects in the operad $\cW{\rm Cob}_{1,{\rm Mor}_\cC}$ and the
       operad composition.
    \label{1CobFig}}
    \end{center}
\end{figure}

We can visualize graphically the spliced colored operad of cobordisms $\cW {\rm Cob}_{1,{\rm Mor}_\cC}$ by
considering a sequence of boxes with matching assigned data $\underline{\epsilon}$ at the left/right boundaries, 
where some of the boxes are empty (and can be filled by plugging in a similar sequence) and some are filled
with 1D cobordisms with defects, as in Figure~\ref{1CobFig}. The colors in this operad are pairs of boundary
data $(\underline{\epsilon}, \underline{\epsilon}')$, namely pairs of objects of ${\rm Cob}_{1,{\rm Mor}_\cC}$. 

\begin{defn}\label{splicedefect}
The category ${\rm Cob}_{1,\cW\cC}$ of 1D cobordisms with spliced defects has objects $\underline{\epsilon}$
as in ${\rm Cob}_{1,{\rm Mor}_\cC}$ and the same cup, cap, and half-lines generators of morphisms, but
in addition to line with a defect marked by a morphism $\phi \in {\rm Mor}_\cC$ there are also generators given
by a line with a defect marked by a sequence of spliced arrows 
$\psi_0 \, \square \, \psi_1 \, \square \, \cdots \, \square \, \psi_n$ in $\cW\cC$.
\end{defn}

\begin{figure}
    \begin{center}
    \includegraphics[scale=0.42]{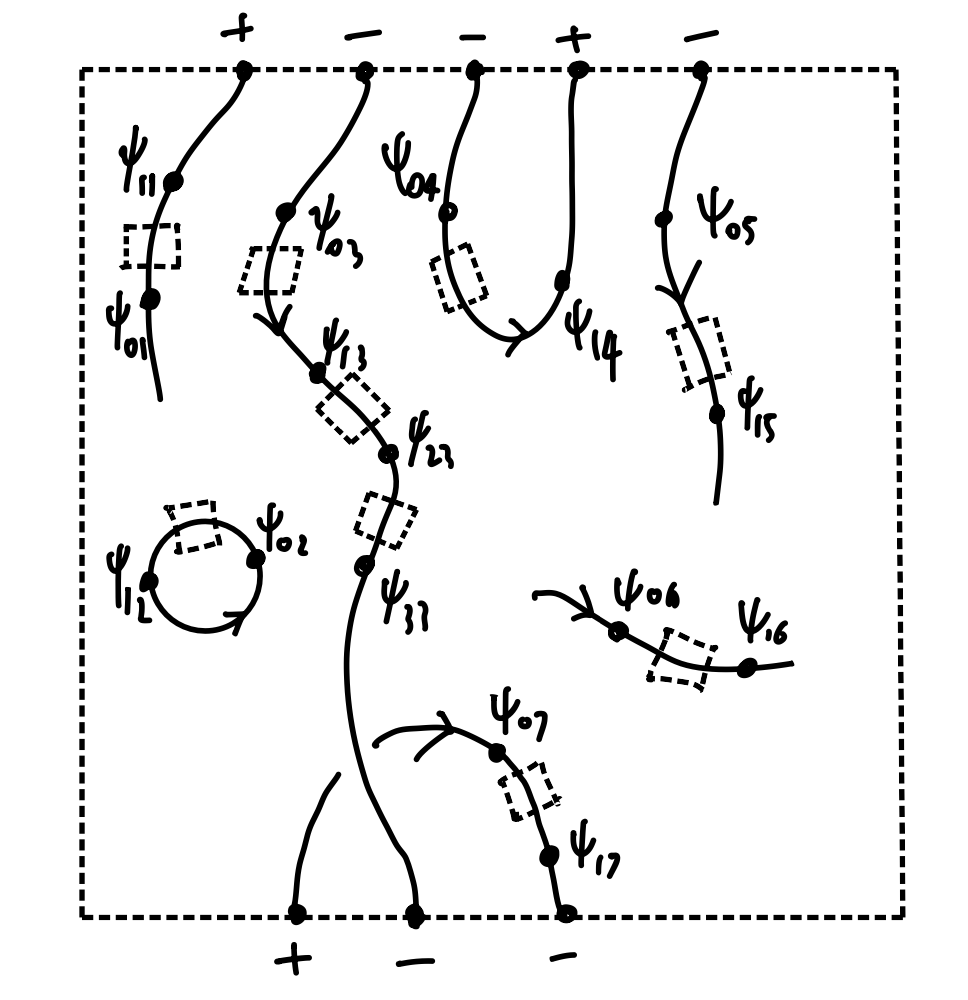}  
    \includegraphics[scale=0.42]{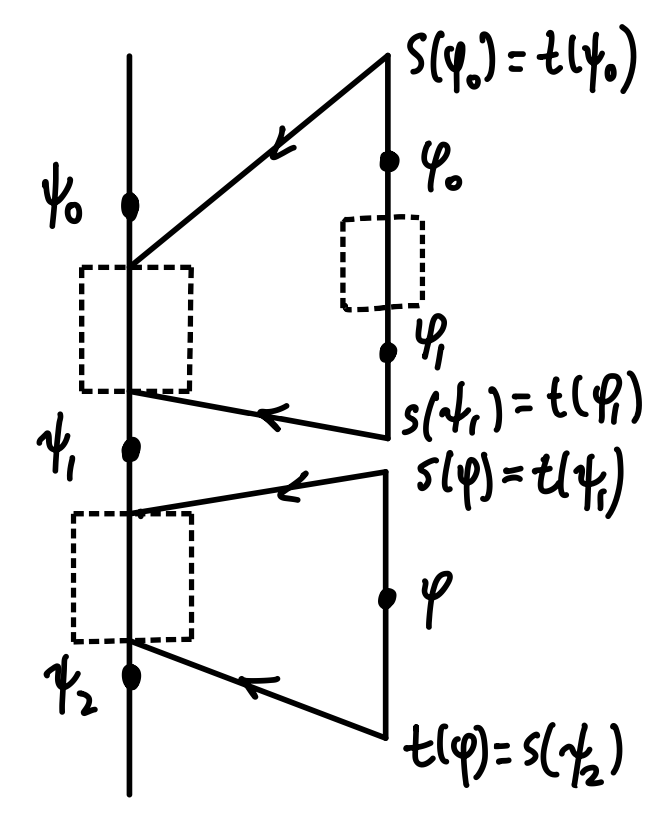} 
       \caption{1D cobordisms with defects in the $\cW\cC$ and the composition of defects.
    \label{2CobFig}}
    \end{center}
\end{figure}

Thus, defects in these cobordisms how have an operadic composition, 
coming from the colored operad of spliced arrows $\cW\cC$. When we form the
corresponding $\cW {\rm Cob}_{1,\cW\cC}$, this has then {\em two} compositional structures, 
one as depicted in Figure~\ref{1CobFig} that composes the cobordisms, and one that composes 
the defects, as in Figure~\ref{2CobFig}.

\begin{lem}\label{OperWCobWC}
The colored operad $\cO_{{\rm Cob},\cC}$ of cobordisms with defects of Definition~\ref{OpCobCdef}
is generated by the operations in $\cW{\rm Cob}_{1,\cW\cC}$. 
\end{lem}

\begin{figure}
    \begin{center}
    \includegraphics[scale=0.42]{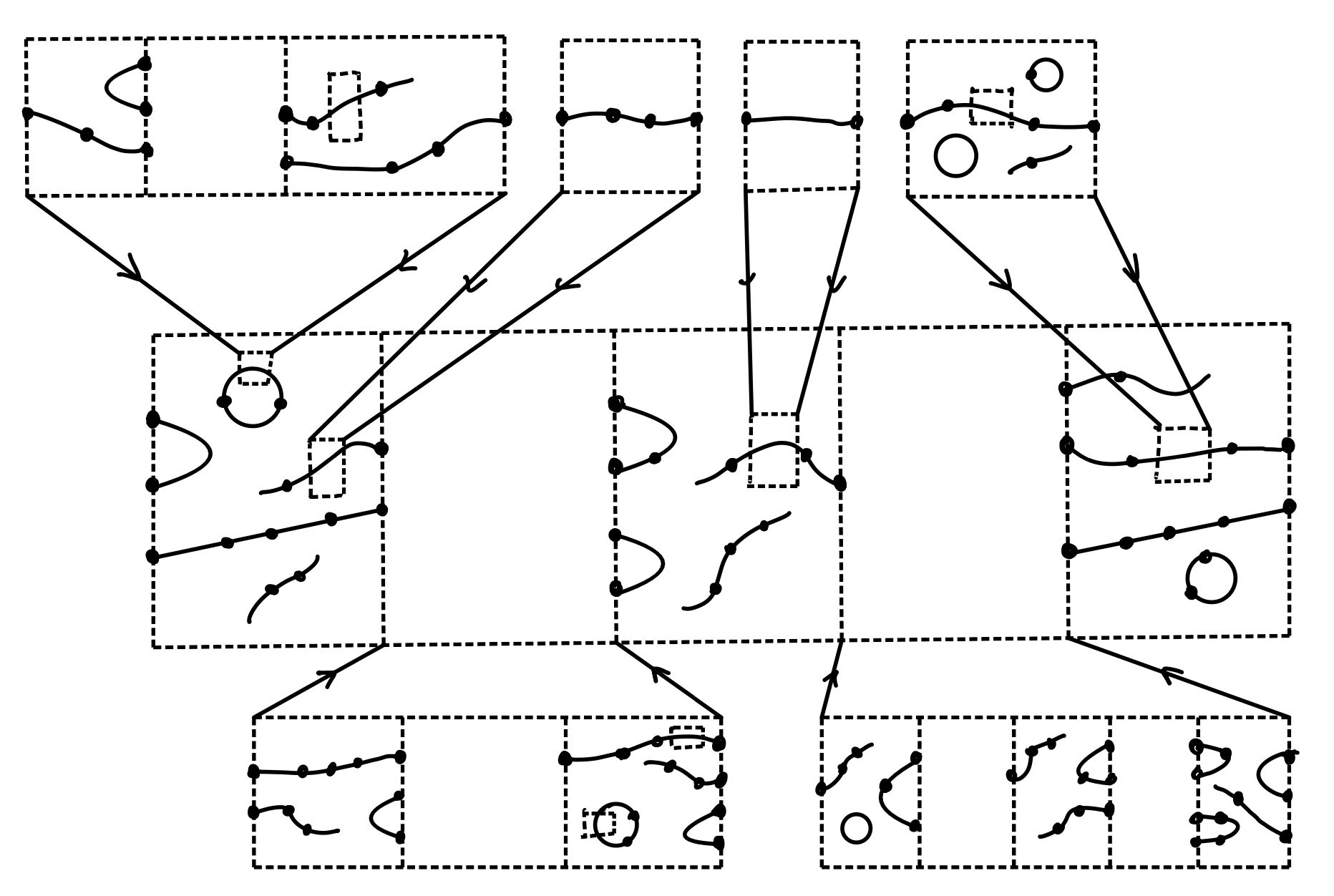}  
       \caption{The operad composition in $\cW {\rm Cob}_{1,\cW\cC}$.
    \label{3CobFig}}
    \end{center}
\end{figure}

\proof Starting with cobordisms in $\cW{\rm Cob}_{1,{\rm Mor}_\cC}$ we can form all the
composition operations, including those of the  two types described above, 
illustrated in Figures~\ref{1CobFig} and \ref{2CobFig} 
combine, and also compositions obtained by filling each box along a line with defects in 
a cobordism of $\cW{\rm Cob}_{1,{\rm Mor}_\cC}$ with any spliced cobordism in 
$\cW {\rm Cob}_{1,\cW\cC}$ with output colors $(+,+)$ or $(-,-)$, not just by a line with
defects in $\cW\cC$. This results in compositions as illustrated in 
Figure~\ref{3CobFig}. The cobordisms obtained as results of these compositions 
are exactly the operations of the operad $\cQ_{{\rm Cob},\cC}$ of Definition~\ref{OpCobCdef}.
\endproof

\begin{thm}\label{WPhiM}
The construction $\cM \mapsto \Phi_\cM$ of a 1D Boolean TQFT with defects from a categorical FSA,
as in Theorem~\ref{catM1DTQFT}, determines a construction 
$\cM \mapsto \Psi_\cM$ of an
operadic Boolean 1D TQFT with defects $\Xi_\cM: \cO_{{\rm Cob},\cC} \to \cO_{R{\rm Mod}}$. 
\end{thm}

\begin{figure}
    \begin{center}
    \includegraphics[scale=0.52]{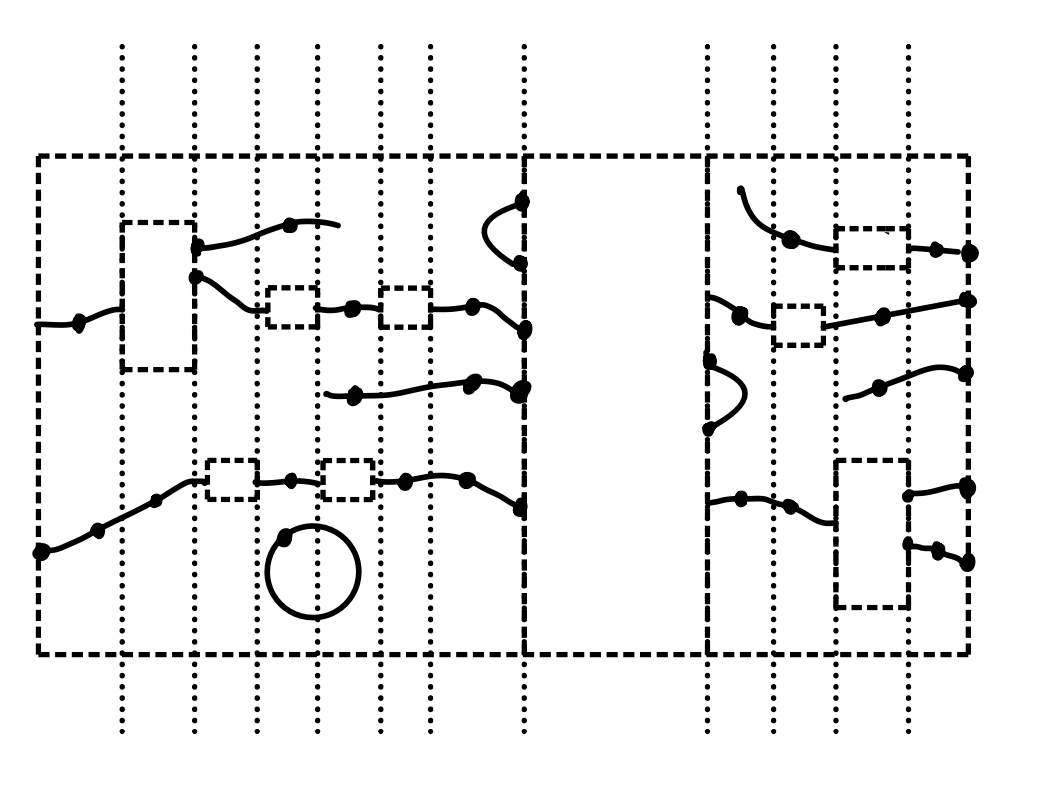}  
       \caption{Decomposition of cobordisms with defect.
    \label{DecompCobordFig}}
    \end{center}
\end{figure}

\proof  First observe that the assignment of the operad of spliced arrows $\cW\cC$ to a category $\cC$
is functorial, so given the fuctor $\tau: \cQ \to \cC$ we obtain a morphism of colored operads $\cW\tau: \cW\cQ \to \cW\cC$. 
Consider the assignment $\cM \mapsto \Phi_{\cM}$  of the corresponding 1D Boolean TQFT with
defects as in Theorem~\ref{catM1DTQFT}. To the functor
$$\Phi_\cM\in {\rm Func}({\rm Cob}_{1,{\rm Mor}_\cC}, \bB\text{-Mod})$$
we can also associate a corresponding morphism of colored operads
$$ \cW\Phi_\cM: \cW{\rm Cob}_{1,{\rm Mor}_\cC} \to \cW\bB\text{-Mod} $$
between the respective colored operads of spliced arrows. 
Given $\cM=(\cQ,\cC, \tau: \cQ\to \cC, (q_0,q_f))$ and $\Phi_\cM$ constructed as in Theorem~\ref{catM1DTQFT}, 
the associated morphism of operads $\cW\Phi_\cM$
assigns to the colors $(\underline{\epsilon}, \underline{\epsilon}')$ corresponding colors in the 
$\cW\bB\text{-{\rm Mod}}$ operads, given by the pairs of modules $(\Phi_\cM(\underline{\epsilon}), 
\Phi_\cM(\underline{\epsilon}'))$ and assigns to a sequence of spliced 
arrows $\fc_0 \, \square \, \fc_1 \, \square \, \cdots \, \square \, \fc_n$ in
the operad $\cW {\rm Cob}_{1,{\rm Mor}_\cC}$ a sequence 
$\Phi_\cM(\fc_0)\, \square \, \Phi_\cM(\fc_1) \, \square \, \cdots \, \square \, \Phi_\cM(\fc_n)$ in 
the operad $\cW\bB\text{-{\rm Mod}}$. 
We write $\cW\cM=(\cW\cQ,\cW\cC, \cW\tau)$ for the morphism of colored operads associated 
to the functor $\tau: \cQ \to \cC$. Passing from $\tau: \cQ\to \cC$ to $\cW\tau: \cW\cQ \to \cW\cC$
in this construction extends cobordisms in ${\rm Cob}_{1,{\rm Mor}_\cC}$ to cobordisms in 
${\rm Cob}_{1,\cW\cC}$. Moreover,  $\cW\tau: \cW\cQ \to \cW\cC$ then determines a functor
$\Psi_{\cW\tau}: {\rm Cob}_{1,\cW\cC} \to \bB\text{-{\rm Mod}}$ that maps objects
$\underline{\epsilon}$ to the same $\bB$-semimodules $\Psi_\cM(\underline{\epsilon})$ and
maps the cup, cap, half-line cobordism to their same images under $\Psi_\cM$, while a line
$\fc=\ell_{\psi_0\square \cdots \square \psi_n}$
marked with a defect $\psi_0\square \cdots \square \psi_n$ is mapped by $\Psi_{\cW\tau}$ to
the homomorphism $T_{\psi_0\square \cdots \square \psi_n}$ that acts as
\begin{equation}\label{Tpsisquare}
 T_{\psi_0\square \cdots \square \psi_n}\, \delta_q = \sum_{q'} \sum_{w_0\square\cdots \square w_n} 
\delta_{s(w_0),q} \, \delta_{t(w_n),q'} \, \delta_{\cW\tau(w_0\square\cdots \square w_n)=\psi_0\square \cdots \square \psi_n} \,\,\, \delta_{q'}\, . 
\end{equation}
This in turn lifts to a morphism $\cW\Psi_{\cW\cQ}: \cW{\rm Cob}_{1,\cW\cC} \to \cW\bB\text{-{\rm Mod}}$, as above. 
To obtain from this a morphism of colored operads $\Xi_\cM: \cO_{{\rm Cob},\cC} \to \cO_{R{\rm Mod}}$, we decompose
a cobordism with defects $\fc$ in $\cO_{{\rm Cob},\cC}(n)$ into a composition of cobordisms $\fc=\fc_k \circ \cdots \circ \fc_1$
where each $\fc_i$ can be factored as a number of lines with defects 
$\sqcup_{i\in \cI} \ell_{\psi_{i,0}\square \cdots \square \psi_{i,n_i}} \sqcup_j \fc_{i,j}$ 
and a number of cobordisms $\fc_{i,j}$ that do not involve defects, see for example Figure~\ref{DecompCobordFig}. 
Then under $\Psi_{\cW\tau}$ these cobordisms will map to
$\Psi_{\cW\tau}(\fc_i)=\otimes_{i\in \cI} T_{\psi_{i,0}\square \cdots \square \psi_{i,n_i}} \otimes_j \Psi_{\cW\tau}(\fc_{i,j})$ 
which is an operation in the operad $\cO_{R{\rm Mod}}$. This assignment is compatible with operad compositions on
both sides, as an operad composition in $\cO_{{\rm Cob},\cC}$ that fills in some of the boxes of $\fc$ with other
cobordisms with defects in $\cO_{{\rm Cob},\cC}$ leads to a refinement of the decomposition of $\fc$ in the form
above, and the corresponding image of the pieces of the decomposition then corresponds to the operad 
composition in $\cO_{R{\rm Mod}}$. 
\endproof

We can also rewrite the action \eqref{Tpsisquare} of the operators $T_{\psi_0\square \cdots \square \psi_n}$
in the following way that will be useful in the following. 

\smallskip

The operad $\cW\cQ$ has an underlying species $\bS_{\cW\cQ}$ that consists of the corollas
for the form
$$ \bT_{\underline{w}} = \Tree[ .$(q_0,q_n')$ [.$\underline{w}$ $(q_0',q_1)$ $(q_1',q_2)$ $\cdots$ $(q'_{n-1},q_n)$ ] ] $$
where the list of morphisms $\underline{w}=(w_0,w_1,\ldots,w_n)$ has sources and targets 
$s(w_i)=q_i$ and $t(w_i)=q_i'$. In fact, given a generating set $X$ of morphisms in the category $\cQ$ 
the species $\bS_{\cW\cQ}$ can be taken with all $w_i \in X$. The operations in $\cW\cQ(n)$ then
consist of all trees $\bT$ with $n$ leaves and vertices given by operations in $\bS_{\cW\cQ}$. For such
$\bT=w_0\square \cdots \square w_n$ we write $s(\bT)=s(w_0)$ and $t(\bT)=t(w_n)$.
We can then rewrite the action \eqref{Tpsisquare}  in the equivalent form
\begin{equation}\label{Tpsisquare2}
 T_{\psi_0\square \cdots \square \psi_n}\, \delta_q = \sum_{q'} \sum_{\bT\in \cW\cQ} \, \delta_{s(\bT),q} \, \delta_{t(\bT),q'}  \delta_{\cW\tau(\bT)=\psi_0\square \cdots \square \psi_n}  \,\,\, \delta_{q'} \, . 
\end{equation}

\subsection{Categorical rational transductions} 

In the usual formulation of context-free languages, a rational transduction $(\cL_{reg}, \fA, \alpha,\beta)$,
with $\cL_{reg}\subset \fA^*$ a regular language and $\alpha: \fA\to \fB$ and $\beta: \fA \to \fC$ monoid
homomorphisms, acts on a context-free language $\cL\subset \fB^*$ by 
$\cL \mapsto \beta(\alpha^{-1}(\cL)\cap \cL_{reg})$. The image of this correspondence is again
a context-free language as all these operations of inverse and direct image and intersection with a
regular language preserve the context free property. The categories of formal languages constructed
in \cite{FerMar} has context-free languages as objects and rational transductions as morphisms,
or PSA automata as objects and rational transducers computing rational transductions as morphisms,
as we recalled in \S \ref{AutoCatSec} of this paper. 

In the categorical version of regular and context-free languages, where context-free languages
$\cL\subset \fB^*$ are replaced more generally by categorical context free grammars described as
operad homomorphisms $P: \cO_\bS \to \cW\cC$,
the action of transducers on context-free languages is formulated in the following way (see  \cite{MelZei},
 \cite{MelZei2}).
 
 Consider a categorical transducer as in Definition~\ref{CatTransd}, and the induced diagram of colored operads of
 spliced arrows
 $$ \xymatrix{ \cQ_T \ar[r]^{\beta} \ar[d]^{\alpha} & \cC' \\
 \cC & } \ \ \ \Rightarrow \ \ \  \xymatrix{ \cW\cQ_T \ar[r]^{\cW\beta} \ar[d]^{\cW\alpha} & \cW\cC' \\ \cW\cC & } $$
 The action of the categorical transducer on categorical context-free grammars is then given by the
 pullback diagram 
 \begin{equation}\label{transactOper}
  \xymatrix{ \cO_\bS\times_{\cW\cC,P,\cW\tau,} \cW\cQ_T \ar[r]^{\quad \pi} \ar[d] & 
 \cW\cQ_T \ar[r]^{\cW\beta} \ar[d]^{\cW\alpha} & \cW\cC' \\
 \cO_\bS \ar[r]^{P} & \cW\cC & } 
\end{equation} 
 so that the categorical transducer maps $P: \cO_\bS \to \cW\cC$ to a new categorical context-free
 grammar $\cW\beta\circ \pi: \cO_\bS\times_{\cW\cC,P,\cW\tau,} \cW\cQ_T \to \cW\cC'$.  As
 shown in  \cite{MelZei2}, this construction exactly correspond to the preimage, intersection with 
 a regular language, and direct image of the usual action of transducers, and all these operations
 preserve the context-free property in the categorical setting as well, in the sense that the resulting
 morphism of operads can be equivalently described as a morphism
 $P': \cO_{\bS'} \to \cW\cC'$ for a resulting species $\bS'$.
 
 The same argument given in Theorem~\ref{CatCatMT} then shows the following.
 
 \begin{prop}\label{CatCFcat}
 The categorical context-free grammars $P: \cO_\bS \to \cW\cC$ are the objects of a category
 with morphisms given by the categorical transducers 
$(\cW\cQ_T, \cW\cC, (\cW\alpha, \cW\beta))$, acting as in \eqref{transactOper}.
 \end{prop}
 
 \smallskip
 \subsection{Categorical context-free grammars and cobordisms with defects}
 
 Consider then a categorical context-free grammar given, as above, by a morphism
 of operads $P: \cO_\bS \to \cW\cC$ as in \cite{MelZei}, \cite{MelZei2}. Let ${\rm Cob}_{1,\cW\cC}$
 be the category of 1D cobordisms with defects labelled by strings in $\cW\cC$, as above.
 Recall from  \cite{MelZei2} that a morphism of operads $P: \cO_\bS\to \cW\cC$ is $\cC$-chromatic
 if it is injective on colors, so that the colors of the species $\bS$ can be seen as a subset of pairs of
 objects $(C,C')$ in $\cC$. 
 
 \begin{prop}\label{OSPWCob}
 A categorical context-free grammar $\cG=(\cC,\bS,S=(C,C'),P: \cO_\bS\to \cW\cC)$ with $\cC$-chromatic $P$ 
 determines a functor  $\Psi_\cG: {\rm Cob}_{1,\cW\cC}\to \bB\text{-{\rm Mod}}$, which in turn induces a morphism
 of operads $\Xi_\cG: \cO_{{\rm Cob},\cC} \to \cO_{\bB{\rm Mod}}$. 
 \end{prop}
 
 \proof The $\cC$-chromatic hypothesis on $P$ implies that the set ${\rm Col}(\bS)$ of colors of $\bS$
 is a subset ${\rm Col}(\bS) \subset {\rm Obj}(\cC)\times {\rm Obj}(\cC)$. Let $\pi_i: {\rm Obj}(\cC)\times {\rm Obj}(\cC)\to
 {\rm Obj}(\cC)$, $i=1,2$, denote the two projections.  Let $\cX_{\cC,\bS}\subset {\rm Obj}(\cC)$ be the finite set defined by
 $$ \cX_{\cC,\bS}=\pi_1({\rm Col}(\bS))\cap \pi_2({\rm Col}(\bS)) \, . $$
 The functor
$\Psi_\cG$ assigns to objects $\underline{\epsilon}$ in ${\rm Cob}_{1,\cW\cC}$ the $\bB$-semimodule
 $\bB\cX_{\cC,\bS}^{\underline{\epsilon}}:=\bB\cX_{\cC,\bS}^{\epsilon_1}\otimes\cdots 
 \otimes \bB\cX_{\cC,\bS}^{\epsilon_n}$, with the same notation previously used. The functor
$\Psi_\cG$ also assign
 to the cap, cup, and half-lines cobordisms the same morphisms of $\bB$-semimodules as
 for the case of FSAs where the distinguished choice of a color $S=(C,C')$ in the datum of the
 categorical context-free grammar provides the initial and final states 
 for the half-line morphisms. The lines labelled by defects in $\cW\cC$ are mapped to
 operators
 \begin{equation}\label{TpsiOSWC}
 T_{\psi_0\square \cdots \square \psi_n}\, \delta_C = \sum_{C' \in \cX_{\cC,\bS}} \, \sum_{\bT\in \cO_\bS} \, 
 \delta_{s(\bT),C} \, \delta_{t(\bT),C'}  \delta_{P(\bT)=\psi_0\square \cdots \square \psi_n}  \,\,\, \delta_{C'} \, .
\end{equation}
Note that the morphism of operads $P: \cO_\bS\to \cW\cC$
has the role here of restricting the defects labelled by operations of $\cW\cC$ to only those operations
that come from operations in $\cO_\bS$. The morphism of operads $\Xi_\cG: \cO_{{\rm Cob},\cC} \to \cO_{\bB{\rm Mod}}$
is then obtained from $\Psi_\cG$ as in the case discussed in Theorem~\ref{WPhiM}. 
\endproof

\begin{thm}\label{TransdactionOSWC}
Consider the functors $\cW\alpha^*\Psi_\cG$ and $\cW\beta^* \Psi_{\cG'}$, seen as functors from
${\rm Cob}_{1,\cW\cQ_T}$ to the category ${\rm Span}(\bB\text{-{\rm Mod}})$, with the modified pullbacks
as in Definition~\ref{pullback2}. Then
the action of transducers $\cT=(\cQ_T,\cC,(\alpha,\beta))$ on categorical context-free grammars
induces natural transformations $\gamma_T :  \cW\alpha^*\Psi_\cG \to \cW\beta^* \Psi_{\cG'}$.
\end{thm}

\proof
First observe that $\cW\alpha: \cW\cQ_T \to \cW\cC$ and $\cW\beta: \cW\cQ_T \to \cW\cC'$
determine functors $\cW\alpha^*: {\rm Cob}_{1,\cW\cQ_T} \to {\rm Cob}_{1,\cW\cC}$ and
$\cW\beta^*: {\rm Cob}_{1,\cW\cQ_T} \to {\rm Cob}_{1,\cW\cC'}$, by the same argument
previously used for the functors $\alpha,\beta$ and the categories of cobordisms with defects
labelled by morphisms (rather than spliced morphisms). 
As before we assume that the categorical transducer has the property that the functor $\alpha$ is
injective on objects, and we also assume the $\cC$-chromatic property at the level of
operads for $P: \cO_\bS \to \cW\cC$. 

In the pullback diagram \eqref{transactOper} we have ${\rm Col}(\bS')={\rm Col}(\bS)\times_{{\rm Col}(\cW\cC)}{\rm Col}(\cW\cQ_T)$. Since $\alpha: \cQ_T \to \cC$ is injective-on-objects and $P: \cO_{\bS}\to \cW\cP$ is $\cC$-chromatic,
the maps ${\rm Col}(\bS)\to {\rm Col}(\cW\cC)={\rm Obj}(\cC)\times {\rm Obj}(\cC)$ and
${\rm Col}(\cW\cQ_T)={\rm Obj}(\cQ_T)\times {\rm Obj}(\cQ_T) \to {\rm Col}(\cW\cC)={\rm Obj}(\cC)\times {\rm Obj}(\cC)$ 
are injective, hence ${\rm Col}(\bS')={\rm Col}(\bS)\cap {\rm Col}(\cW\cQ_T)$ and $\cX_{\cQ_T,\bS'}=\cX_{\cC,\bS}\cap {\rm Obj}(\cQ_T)$. We have $\Psi_\cG(+)=\bB\cX_{\cC,\bS}$. Similarly, we have $\Psi_{\cG'}(+)=\bB\cX_{\cC',\bS'}$. The functor
$\beta: \cQ_T \to \cC'$ is not necessarily assumed to be injective-on-objects, but it is finitary, so the preimage
$\beta^{-1}(C')$ for any $C'\in {\rm Obj}(\cC')$ is a finite set. Thus, we have a finite-to-one projection
$\cX_{\cQ_T,\bS'} \to \cX_{\cC',\bS'}$. This induces a corresponding projection $\bB\cX_{\cQ_T,\bS'}\to \bB\cX_{\cC',\bS'}$.
The modified pullback $\cW\alpha^*\Psi_\cG$ has  
$\cW\alpha^*\Psi_\cG(+)=\Psi_\cG(+)_{\cW\alpha}=\Pi_{\cW\alpha}\Psi_\cG(+)$, 
the subspace invariant under all the
$T_{\cW\alpha(w_0\square\cdots\square w_n)}$ acting as
$$ T_{\cW\alpha(w_0\square\cdots \square w_n)}\delta_C=\sum_{\hat C\in \cX_{\cC,\bS}} \,\,
\sum_{\bT\in \cO_{\bS}, s(\bT)=C, t(\bT)=\hat C}\,\,  \delta_{P(\bT)=\cW\alpha(w_0\square\cdots \square w_n)}\,\, \delta_{\hat C}. $$
Up to finite multiplicities that do not count with Boolean coefficients this can be rewritten as
$$ T_{\cW\alpha(w_0\square\cdots \square w_n)}\delta_q=\sum_{\hat q\in \cX_{\cC,\bS}\cap {\rm Obj}(\cQ_T)} \,\,
\sum_{\bT\in \cO_{\bS}, s(\bT)=q, t(\bT)=\hat q}\,\, \delta_{P(\bT)=\cW\alpha(w_0\square\cdots \square w_n)}\,\, \delta_{\hat q} $$
$$ =\sum_{\hat q\in \cX_{\cQ_T,\bS'}} \,\, \sum_{\bT'\in \cO_{\bS'}, s(\bT')=q, t(\bT')=\hat q}\,\, \delta_{\pi(\bT')=w_0\square\cdots \square w_n}\,\, \delta_{\hat q} \, . $$
Thus, we have $\Pi_{\cW\alpha}\Psi_\cG(+)=\bB (\cX_{\cC,\bS}\cap {\rm Obj}(\cQ_T))=\bB\cX_{\cQ_T,\bS'}$.
 
 For the modified pullback $\cW\beta^*\Psi_{\cG'}$ we have 
 $\Psi_{\cG'}(+)=\Psi_{\cG'}(+)_{\cW\beta}=\bB\cX_{\cC',\bS'}$. 
 The $T_{\cW\beta(w_0\square\cdots \square w_n)}$ operators act as
 $$ T_{\cW\beta(w_0\square\cdots \square w_n)} \delta_{C'}
=\sum_{\hat C'\in \cX_{\cC',\bS'}} \,\,
\sum_{\bT'\in \cO_{\bS'}, s(\bT')\in \beta^{-1}(C'), t(\bT')\in \beta^{-1}(\hat C')} \,\, 
\delta_{P'(\bT')=\cW\beta(w_0\square\cdots \square w_n)}\,\, \delta_{\hat C'}. $$
This lifts to an operator
$$ T_{\cW\beta(w_0\square\cdots \square w_n)} \delta_{q'} =\sum_{\hat q'\in \cX_{\cQ_T,\bS'}}
\sum_{\bT'\in \cO_{\bS'}, s(\bT')=q', t(\bT')=\hat q'}
\delta_{\pi(\bT')=w_0\square\cdots \square w_n} \delta_{\hat q'} \, , $$
with coefficients that are constants along the fibers of the
projection $\cX_{\cQ_T,\bS'}\to \cX_{\cC',\cS'}$, since they only depend 
on the fibers $\beta^{-1}(C')$ and $\beta^{-1}(\hat C')$.

Consider then, for an object $\underline{\epsilon}$ in the category of cobordisms,
the $\bB$-semimodule $\otimes_i \bB\cX_{\cQ_T, \bS'}^{\epsilon_i}$, together with 
the homomorphisms of $\bB$-semimodules 
\begin{equation}\label{CorrBX}
 \xymatrix{ & \otimes_i \bB\cX_{\cQ_T, \bS'}^{\epsilon_i}  \ar[dl]^{f_{\underline{\epsilon}}} \ar[dr]_{g_{\underline{\epsilon}}}  & \\
\otimes_i \bB\cX_{\cC,\bS}^{\epsilon_i} & & \otimes_i \bB\cX_{\cC',\bS'}^{\epsilon_i} } 
\end{equation}
where $f_{\underline{\epsilon}}=\otimes_i f_{\epsilon_i}$, with $f_{\epsilon_i}$ 
given by the inclusion of $\bB\cX_{\cC',\bS'}^{\epsilon_i} \subset \bB\cX_{\cC,\bS}^{\epsilon_i}$ 
as the subspace $\Pi_{\cW\alpha}\Psi_\cG(\epsilon_i)$ and $g_{\underline{\epsilon}}=\otimes_i g_{\epsilon_i}$,
with $g_{\epsilon_i}$ the projection 
$\bB \cX_{\cQ_T,\bS'}^{\epsilon_i}\to \bB \cX_{\cC',\cS'}^{\epsilon_i}$. We define the
natural transformation $\gamma_{T,\underline{\epsilon}}$ as this diagram in the category
${\rm Span}(\bB\text{\rm-Mod})$. To see that this is indeed a natural transformation it suffices to
check the commutativity of the relevant diagram in the case $\underline{\epsilon}=+$ as the 
more general case then follows. Consider then the diagram
$$ \xymatrix{ & \bB\cX_{\cQ_T, \bS'}  \ar[dl]_{f} \ar[dr]^{g}  \ar[dd]^{T_{w_0\square\cdots\square w_n}} & \\
\bB\cX_{\cC,\bS} \ar[dd]_{T_{\cW\alpha(w_0\square\cdots\square w_n)}} & & \bB\cX_{\cC',\bS'} \ar[dd]^{T_{\cW\beta(w_0\square\cdots\square w_n)}} \\
& \bB\cX_{\cQ_T, \bS'}  \ar[dl]^{f} \ar[dr]_{g}  & \\
\bB\cX_{\cC,\bS} & & \bB\cX_{\cC',\bS'}  } $$
that describes $\gamma_{T,+}$.
By the commutative
pullback diagram \eqref{transactOper} we can write $\pi_1^*(P): \cO_{\bS'}\to \cW\cC$, for $\cO_{\bS'}=\cO_{\bS} \times_{\cW\cC,P,\cW\alpha} \cW\cQ_T$ and $P: \cO_\bS\to \cW\cC$ equivalently as the composition
$$ \cO_{\bS'} \stackrel{\pi}{\rightarrow} \cW\cQ_T \stackrel{\cW\alpha}{\rightarrow} \cW\cC $$
while the resulting $P': \cO_{\bS'}\to \cW\cC'$ is the composition
$$ \cO_{\bS'} \stackrel{\pi}{\rightarrow} \cW\cQ_T \stackrel{\cW\beta}{\rightarrow} \cW\cC' \, . $$
The morphism of colored operads $\pi: \cO_{\bS'} \to \cW\cQ_T$ determines by Proposition~\ref{OSPWCob}
a functor $\Psi: {\rm Cob}_{1,\cW\cQ_T} \to \bB\text{-{\rm Mod}}$, which maps a line with
defect labelled by $w_0\square\cdots \square w_n$ in $\cW\cQ_T$ to  
\begin{equation}\label{Psiell}
 \Psi(\ell_{w_0\square\cdots \square w_n})\delta_q=T_{w_0\square\cdots \square w_n}\delta_q=\sum_{q' \in \cX_{\cQ_T,\bS'}} \,\,\sum_{\bT'\in \cO_{\bS'}, s(\bT')=q, t(\bT')=q'} \,\, \delta_{\pi(\bT)=w_0\square\cdots \square w_n}\,\, \delta_{q'}\, ,
\end{equation} 
which gives the middle arrow of the diagram. This map only differs by finite multiplicities (which are invisible with Boolean
coefficients) from 
$$ T_{\cW\alpha(w_0\square\cdots \square w_n)} \delta_q =\sum_{q'} \sum_{\bT\in \cO_{\bS}, s(\bT)=q, t(\bT)=q'} \,\,
\delta_{P(\bT)=\cW\alpha(w_0\square\cdots \square w_n)}\,\, \delta_{q'}\, ,$$ 
so that the left-part of the diagram commutes. In the right-side of the diagram, we compare the action of \eqref{Psiell}
on a basis element $\delta_q$ with the action of $T_{\cW\beta(w_0\square\cdots \square w_n)}$ 
on a basis element $\delta_{C'}$. As we have seen, we can write \eqref{Psiell} as
$$ T_{\cW\alpha(w_0\square\cdots \square w_n)} \sum_{q\in \beta^{-1}(C')} \delta_q= 
\sum_{q\in \beta^{-1}(C')} \sum_{q' \in \beta^{-1}(hat C')} \sum_{\bT'\in \cO_{\bS'}, s(\bT')\in \beta^{-1}(C), t(\bT')=\beta^{-1}(\hat C'} 
\delta_{P'(\bT)=\cW\beta(w_0\square\cdots \square w_n)}\,\, \delta_{q'}\, , $$
which agrees with $T_{\cW\alpha(w_0\square\cdots \square w_n)} \delta_{C'}$. Thus, the right-side of the
diagram also commutes. 
The commutativity for morphisms induced by other cobordisms is also satisfied. For the cup cobordism,
since the coefficients are Boolean, we have $\sum_q q^\vee(q)=1$ regardless of restricting to a subspace
or projecting to a quotient space. For the cap cobordism, on the left-side we have
$1 \mapsto \sum_q \Pi_\alpha q \otimes \Pi_\alpha q^*$ (by the definition of the modified pullback on
morphisms) which matches the arrow in the middle, and on the right-side under the projection we
have $1\mapsto \sum_C C\otimes C^\vee$ which differs from projecting the image of the middle arrow
only by multiplicities of the fibers of $\cX_{\cQ_T,\bS'}\to \cX_{\cC',\cS'}$, but these multiplicities
do not appear in the Boolean sum. The case of the half-arrows also matches as the intial and final state
are mapped to initial and final state by the correspondence.
\endproof

\subsection{Operads and Chomsky--Sch\"utzenberger representation}

It is shown in \cite{MelZei2} that the functorial mapping $\cC \to \cW\cC$ of
categories to their operads of spliced arrows has a left-adjoint functor, which
we denote here by ${\rm Cont}$ that maps an operad $\cO$ to its {\em contour
category} ${\rm Cont}(\cO)$. An explicit description of the contour category
in terms of generating arrows that follow the contours of trees in the operad $\cO$
is described in \S 3.2 of \cite{MelZei2} and we will not recall it here in detail.
We only mention the fact that in the case of a free operad $\cO_\bS$
over a species $\bS$, the contour category ${\rm Cont}(\cO_\bS)$ 
is the category with objects $N^d$, $N^u$ for each color $R$ of $\bS$, and with 
morphisms freely generated by arrows $(x,i): N_i^d \to N_{i+1}^u$, with $i=0,\ldots,n$, 
for every node of the form
$$ \Tree[.$N_0$ [.$x$ $N_1$ $N_2$ $\cdots$ $N_n$ ]] $$
in the species $\bS$. For a given species $\bS$ with a chosen color $S$, the
tree-contour categorical context-free grammar 
$$ {\rm TreeCont}_{\bS,S} =({\rm Cont}(\cO_\bS), \bS, S, P_\bS: \cO_\bS \to \cW{\rm Cont}(\cO_\bS)) \, , $$
where $P_\bS: \cO_\bS \to \cW{\rm Cont}(\cO_\bS)$ is the unit of the adjunction of the $\cW$ and ${\rm Cont}$
functors. (This is referred to in  \cite{MelZei2} as the ``universal context-free grammar".) The collection
of the ${\rm TreeCont}_{\bS,S}$ for data $(\bS,S)$ gives a categorical version of the 
Dyck languages of the Chomsky--Sch\"utzengerger representation theorem of \cite{ChoSchu}.

\begin{prop}\label{DyckTQFTs}
The assignment of Proposition~\ref{OSPWCob} of functors 
$\Psi_\cG: {\rm Cob}_{1,\cW\cC}\to {\rm Span}(\bB\text{-{\rm Mod}})$ to
categorical context-free grammars $\cG=(\cC,\bS,S=(C,C'),P: \cO_\bS\to \cW\cC)$ 
with $\cC$-chromatic $P$ 
is completely determined by the values on $\cC$-chromatic tree-contour grammars
and by the functoriality under transducers proved in Theorem~\ref{TransdactionOSWC}.
\end{prop}

\proof
It is proven in \cite{MelZei2} that any categorical context-free grammar 
$\cG=(\cC, \bS, S, P: \cO_\bS \to \cW\cC)$ factors through the tree-contour ${\rm TreeCont}_{\bS,S}$
with the same $(\bS,S)$. Namely, there is a functor $\tau_\cG: {\rm Cont}(\cO_\bS) \to \cC$ and
that gives rise to a commutative diagram of operad homomorphisms
$$ \xymatrix{ \cO_\bS \ar[rr]^{P} \ar[dr]^{P_\bS} & & \cW\cC \\ 
& \cW{\rm Cont}(\cO_\bS) \ar[ur]^{\cW\tau_\cG} & } $$
Moroever, any categorical context-free grammar 
$\cG=(\cC, \bS, S, P: \cO_\bS \to \cW\cC)$ is obtained as
the image of a $\cC$-chromatic tree-contour grammar under a transducer, in the following way,
as shown in  \cite{MelZei2}.
Given $P: \cO_\bS \to \cW\cC$, the underlying map of species $\phi: \bS \to \bS_{\cW\cC}$ factors
as $\phi=\phi_V\circ \phi_{\rm Col}$, with $\phi_{\rm Col}: \bS \to \phi_{\rm Col}\bS$ that acts
as the identity on vertices and is surjective on colors and $\phi_V: \phi_{\rm Col}\bS\to \bS_{\cW\cC}$
that acts like $\phi$ on vertices but is injective on colors, so that the resulting
$P_V: \cO_{\phi_{\rm Col}\bS} \to \cW\cC$ is $\cC$-chromatic. By the previous universality
theorem, there is a functor $\tau_V: {\rm Cont}(\cO_{\phi_{\rm Col}\bS})\to \cC$, so that we
obtain a diagram 
$$ \xymatrix{ \cO_{\phi_{\rm Col}\bS} \ar[rr]^{P_V} \ar[dr]^{P_{\phi_{\rm Col}\bS}} & &  \cW\cC \\
& \cW\cO_{\phi_{\rm Col}\bS}  \ar[ur]^{\cW\tau_V} &  } $$
So that one obtains $\cG$ through a diagram of the form
$$ \xymatrix{ \cO_\bS \ar[r]^{P_\bS} \ar[d]_{\phi_{\rm Col}} & \cW{\rm Cont}(\cO_\bS) 
\ar[r]^{\cW\tau_\cG} \ar[d]^{\cW\phi_{\rm Col}} & \cW\cC \\
\cO_{\phi_{\rm Col}\bS} \ar[r]^{P_{\phi_{\rm Col}\bS}\quad\quad} & \cW{\rm Cont}(\cO_{\phi_{\rm Col}\bS})
 & } $$
Thus, this categorical version of the Chomsky--Sch\"utzengerger representation theorem shows
that the construction of the 1D TQFT's with defects associated to categorical context-free grammars
are entirely determined by the ones associated to the $\cC$-chromatic tree-contour grammars 
and the action of transducers by natural transformations.
\endproof 

\medskip
\subsection{Further directions: higher dimensional automata and TQFTs with defects} 

There is a notion of higher dimensional automata (see \cite{GhaKurz}), based on James 
Roger's ``higher dimensional trees" \cite{Rogers}, and allowing for a rich
categorical formulation and for algebra and coalgebra structures. It seems then
natural to investigate the
question of possible relations between these higher dimensional automata and
TQFTs with defects, especially in the 2-dimensional case where ordinary TQFTs
admit an algebraic description in terms of Frobenius algebras, and versions with
defects have been analyzed, for instance, in \cite{Carq}. We will return to this
question in future work.

\end{document}